\documentclass{aa} 
\usepackage{graphicx}
\usepackage{gensymb}
\usepackage{placeins}

\usepackage{txfonts}

\usepackage[hidelinks]{hyperref}
\hypersetup{
    colorlinks=true,
    allcolors=blue,}

\begin{document}

    \title{AGN - host galaxy photometric decomposition using a fast, accurate and precise deep learning approach}
    \titlerunning{AGN - host galaxy photometric decomposition with deep learning}

    \subtitle{}

    \author{B. Margalef-Bentabol\inst{1}\thanks{\email{B.Margalef.Bentabol@sron.nl}}\and 
          L. Wang\inst{1, 2}\and 
          N. Pandolfo\inst{4}\and
          A. La Marca\inst{1, 2}\and
          V. Rodriguez-Gomez\inst{3}\and
          Y. Fu\inst{2,5}\and
          M. Chen\inst{2}
          }

   \institute{SRON Netherlands Institute for Space Research, Landleven 12, 9747 AD Groningen, The Netherlands
   \and
   Kapteyn Astronomical Institute, University of Groningen, Postbus 800, 9700 AV Groningen, The Netherlands
   \and
   Instituto de Radioastronom\'ia y Astrof\'isica, Universidad Nacional Aut\'onoma de M\'exico, Apdo. Postal 72-3, 58089 Morelia, Mexico
   \and
    ??
   \and
   Leiden Observatory, Leiden University, Einsteinweg 55 2333 CC Leiden, The Netherlands805}

   \date{Received -; accepted -}

 
 \abstract
  {}
  {Identification of active galactic nuclei (AGN) is extremely important for understanding galaxy evolution and its connection with the assembly of supermassive black holes (SMBH). With the advent of deep and high angular resolution imaging surveys such as those conducted with the {\it James Webb} Space Telescope ({\it JWST}), it is now possible to identify galaxies with a central point source out to the very early Universe. In this proof of concept study, we aim to develop a fast, accurate and precise method to identify galaxies which host AGN and recover the intrinsic AGN contribution fraction ($f_{AGN}$).}
  {We trained a deep learning (DL) based method Zoobot to estimate the fractional contribution of a central point source to the total light. Our training sample comprises realistic mock \textit{JWST} images of simulated galaxies from the IllustrisTNG cosmological hydrodynamical simulations. We injected different amounts of the observed \textit{JWST} point spread functions to represent galaxies with varying levels of AGN contribution. Galaxies in our training sample span a wide range of morphologies, including mergers. We analyse in detail the performance of our method as a function of various galaxy properties and compare with results obtained from the traditional light profile fitting tool GALFIT. After training, we applied our method to real \textit{JWST} observations, in the COSMOS field.
  }
  {We find excellent performance of our DL  method in recovering the injected $f_{AGN}$, in terms of precision and accuracy. The mean difference between the predicted and true  $f_{AGN}$ is $-0.002$ and the overall root mean square error (RMSE) is $0.013$.  The overall relative absolute error (RAE) is $0.076$ and the outlier (defined as predictions with RAE $>20\%$) fraction is $6.5\%$. In comparison, using GALFIT, we achieve a mean difference of -$0.02$, RMSE of $0.12$, RAE of $0.19$ and outlier fraction of $19\%$. We also investigate how these key performance metrics obtained from Zoobot and GALFIT vary as a function of the injected $f_{AGN}$, redshift, signal-to-noise ratio, and galaxy size. In addition to the superior performance, our DL method has several other advantages over traditional methods. For example, it has a much higher success rate (even for highly disturbed or irregular galaxies) and is extremely fast. We applied our trained DL model to real {\it JWST} observations and found that 20\% of the X-ray-selected AGN and 8\% of the MIR-selected AGN are also identified as AGN using a cut at $f_{\rm AGN} > 0.2$. When using $f_{\rm AGN} > 0.1$, these overlaps increase to 33\% for the X-ray AGN and 15\% for the MIR AGN. In summary, our DL-based method to identify AGN and estimate AGN contribution fraction has a huge potential in future applications to large galaxy imaging surveys.
  }
  {}

  \maketitle
%

\section{Introduction}

It is generally accepted that there exists co-evolution between supermassive black holes (SMBH) and their host galaxies \citep[][]{Kormendy2013}. One manifestation of this co-evolution is the tight correlation between the mass of an SMBH and its host galaxy properties, such as the stellar velocity dispersion, bulge luminosity and bulge mass \citep{Gultekin2009, Beifiori2012, Graham2013, McConnell2013, Lasker2014}. These correlations may hint at a fundamental connection between the central SMBH and the formation and evolution of its host galaxy. Theoretical models show that SMBH feedback, via radiative heating, outflows, or jets could potentially explain these correlations, by regulating or halting the growth of itself and the host galaxy \cite[e.g.,][]{Somerville2008, Booth2009, Weinberger2018, Dave2019}. However, these correlations could also arise from merging events, in which mergers could explain, for example, the growth of SMBH mass and the stellar mass of the host galaxy \cite[e.g.,][]{Croton2006, Peng2007, Hirschmann2010, Jahnke2011}. To better understand the co-evolution (or not) of SMBHs and host galaxies, it is important to study their link at different cosmic times. At high redshifts, studies have to rely on accreting SMBHs, which are active galactic nuclei (AGN),  to be able to obtain SMBH mass estimates, and in particular Type I AGN. It is also necessary to obtain quantitative measures of the physical properties of the host galaxies. However, the presence of a bright central AGN can make this task very difficult, particularly for the bulge component, as galaxies with significant contribution from the AGN to the total flux will appear more bulge-dominated \citep{Pierce2010}. 

It is crucial to correctly separate the central AGN light from the host galaxy, across a wide range of galaxy types and redshifts. Good quality optical imaging, with high spatial resolution and signal-to-noise ratio (S/N), is needed to decompose the observed total light into contributions from the host galaxy and the central AGN. Traditionally, this is done in photometric data by fitting two-dimensional (2D) profiles to the galaxy's light, using one (or more if needed) analytic profile to describe the galaxy (typically a S\'ersic profile) and a point spread function (PSF) profile to describe the central point source. Many studies use GALFIT \citep{Peng2002}, one of the most widely used 2D surface brightness modelling software to perform image-based decomposition of AGN and host galaxy light \citep[e.g.][]{Kim2008, Bentz2009, Gabor2009, Bohm2013, Schramm2013, Du2014, Urbano2019, Son2022, Aird2022, Ji2022, Dewsnap2023, Zhuang2023, Sturm2024, Zhuang2024}. However, there are some known technical issues with GALFIT. For example, in some cases, the minimisation algorithm used can be trapped in a local minimum, leading to unreliable fits. Other software have been used in an attempt to improve on GALFIT performance, such as PSFMC \citep{Mechtley2014}, which is a Markov Chain Monte Carlo (MCMC) simultaneous fitting software to perform multi-component profile fitting \citep{Mechtley2016, Marshall2021} and LENSTRONOMY \citep{Birrer2015, Birrer2018}, which uses particle swarm optimisation \citep{Kennedy1995} for $\chi^2$ minimisation, to reduce the likelihood of getting trapped in a local minimum when searching the parameter space, and MCMC for Bayesian parameter inference \citep{Foreman-Mackey2013}. The latter method is used by \citep{Li2021} to decompose a sample of X-ray-selected AGN into quasars and host galaxy components, to investigate the properties of the host galaxies. However, these approaches all assume that the galaxy's surface brightness profile can be well fitted by a single S\'ersic profile (or combination of S\'ersic profiles), which may not be the case, particularly for irregular galaxies, highly disturbed merging galaxies, or galaxies with complicated substructures. 

Consequently, if S\'ersic profiles cannot adequately describe the host galaxy light, it can introduce a systematic bias in the derived luminosity of the PSF component and of the host galaxy. In addition, if there is a significant contribution to the total flux from the central AGN, it can lead to a bias in the morphology of the host galaxy, as it may appear more bulge-dominated \citep{Pierce2010}. Indeed, it is not always a good idea to represent a galaxy with a single S\'ersic profile. \citet{Bentz2009} study AGN host galaxies at redshift $z \approx 0.7$, by fitting their surface brightness distributions with a combination of S\'ersic profiles (to account for the host galaxy's light) and PSF profile (to describe the AGN component). They find that most galaxies in their sample are well described with two S\'ersic profiles, in addition to the PSF profile, and a few of them required three or more S\'ersic profiles to describe additional components such as bars. Another possible difficulty in decomposing the AGN from the host galaxy is the (sometimes significant) variation of the PSF in a given survey due to spatial/temporal changes or differences in galaxy spectral energy distributions (SEDs), as an incorrect PSF can bias the estimated contribution of the AGN to the total flux of the galaxy. \citet{Kim2008} perform 2D decomposition of S\'ersic + PSF profiles to galaxies hosting AGN in {\it Hubble Space Telescope} (HST) images and investigate the effect of realistic PSF mismatch, finding a systematic overestimation of the flux of the host galaxies, particularly for those containing bright AGN.

Without additional information on the possible presence of AGN activity in a galaxy (e.g., from X-ray, mid-infrared or radio observations), it is not always easy to discern whether a galaxy has an AGN component in the form of a central point source from 2D surface brightness fitting, particularly for galaxies with very concentrated light profiles. This is because galaxy light in some cases could be more or less equally well described by a single S\'ersic component or a combination of S\'ersic + PSF profiles. A more complex model will always have a smaller $\chi^2$. Therefore, a better fit is not a necessarily good indicator to decide between different models. To mitigate this problem, \citet{Aird2022} fits the {\it HST} imaging of a sample of galaxies in the Cosmic Assembly Near-infrared Deep Extragalactic Legacy Survey \citep[CANDELS;][]{Koekemoer2011, Grogin2011} at $z = $ 0.5 – 3 with a  S\'ersic profile plus an additional central point source component and with a S\'ersic profile only, and then used the residual flux fraction \citep[RFF;][]{Hoyos2011} to determine which is the best model. The RFF measures the fraction of the flux contained in the residual image that cannot be explained by fluctuations in the background. However, good estimates of the background and galaxy's size are needed for this method.

In this work, we present a promising new methodology to determine the AGN contribution to the total flux of a galaxy in imaging data, by combining deep learning (DL) methods and cosmological hydro-dynamical simulations. Over the last decade or so, DL methods have been widely used for diverse astronomy applications \citep{Dieleman2015, Huertas-Company2018, Walmsley2020, Margalef2020, Zanisi2021, Huertas2023}. In particular, they show great success in different image-based astronomical problems, such as morphological classifications of galaxies \citep[e.g.,][]{Huertas2015, Dominguez2018, Cheng2020, Walmsley2022a}, merger identifications \citep{Bottrell2019, Ferreira2020, Ciprijanovic2020, Pearson2022, Bickley2021, Margalef2024} and determining galaxy physical properties and structural parameters \citep{Tuccillo2018, Simet2021, Bisigello2023}. On the other hand, the use of cosmological hydro-dynamical simulations allows us to create a comprehensive training sample of diverse and realistic galaxies, spanning a large range in redshift and mass.

Our DL model is trained on mock images of simulated galaxies with different levels of AGN contributions to the total flux. The construction of our training sample can easily incorporate the full information on the expected variations of the PSF. Therefore, our DL model can learn to infer the intrinsic AGN fraction while automatically folding in the impact of different PSFs. In comparison, while it is possible to examine the goodness-of-fit for different PSF models for methods based on light profile fitting such as GALFIT, in practice it will be extremely time-consuming, particularly for large samples of galaxies. Another advantage of our DL-based method is that it does not rely on an assumed (and often simplified) galaxy surface brightness profile which in many cases is not able to fully describe a galaxy's light profile and thus can introduce biases in the estimation of the AGN contribution. Finally, our method, like any other machine learning-based method, has the advantage of being very fast to implement in new data once the DL model has been trained, making it much more computationally efficient than any traditional method based on light profile fitting. 

This paper is organised as follows. In Section  \ref{sect:data}, we describe the observed {\it James Webb} Space Telescope \citep[{\it JWST};][]{Gardner2006} imaging data used in this work and the generation of the corresponding mock \textit{JWST} images of simulated galaxies selected from the IllustrisTNG. Of particular importance are the real \textit{JWST} PSF models and their variations within the survey data. In addition, we introduce two AGN samples, selected in the X-ray and mid-infrared (MIR), which we use to compare with AGN identified using our DL-based method. In Section \ref{sect:Methods}, first we explain how we create the final images mimicking galaxies containing AGN by injecting different levels of PSF contribution (taken as AGN contribution fractions) in the mock \textit{JWST} images. Then we introduce our DL-based method \citep[Zoobot;][]{Walmsley2023} to recover the intrinsic AGN contribution fraction in the observed total light. To compare with traditional light profile fitting-based methods, we also briefly describe GALFIT and how we use it in this work.
In Section \ref{sect:Results}, we explore in detail the results from both methods and compare their performances as a function of various galaxy properties. We also present the first application of our DL-based method to real \textit{JWST} data and compare with AGN selected in the X-ray and MIR. Finally, in Section  \ref{sect:Conclusions} we summarise the paper and highlight the main conclusions of our work.

Throughout the paper we assume a flat $\Lambda$CDM universe with $\Omega_M=0.2865$, $\Omega_{\Lambda}=0.7135$, and $H_0=69.32$ km s$^{-1}$ Mpc$^{-1}$ \citep{Hinshaw2013}.
Unless otherwise stated, magnitudes are presented in the AB system.

\section{Data}\label{sect:data}

In this section, we first introduce the real \textit{JWST} observations and the PSF models obtained from the COSMOS-Web survey. Then, we describe the synthetic \textit{JWST} images generated using simulated galaxies selected from the IllustrisTNG cosmological hydrodynamical simulations. The observed and simulated datasets are combined later to create mock  \textit{JWST} galaxy images with different levels of AGN contribution.

\subsection{JWST/COSMOS-WEB}\label{subsec.data.jwst}

\begin{figure*}
\centering
   \includegraphics[width=0.33\textwidth]{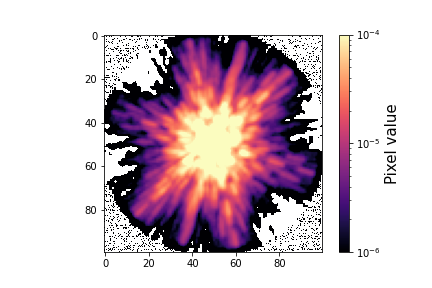}
    \includegraphics[width=0.33\textwidth]{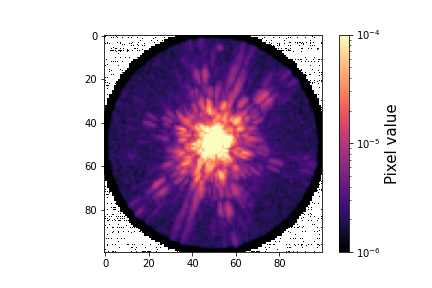}
    \includegraphics[width=0.33\textwidth]{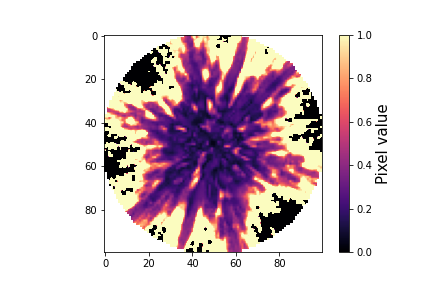}
    \caption{Overview of the \textit{JWST}/NIRCam F150W PSFs in COSMOS-Web. We stacked all available PSFs from \citet{Zhuang2024} and show the mean PSF (left), standard deviation (centre) and the coefficient of variation (right), calculated pixel by pixel. The PSFs have been rebinned to a pixel resolution of $0.03\ \arcsec/\text{pixel}$, matching the resolution used for the synthetic image creation. The axes show the number of pixels, corresponding to 3\arcsec across. The colorbar shows the value of each pixel.}
    \label{fig.psf}
\end{figure*}

COSMOS-Web \citep[][PIs: Kartaltepe \& Casey, ID=1727]{Casey2023} is a 255-hour \textit{JWST} treasury imaging survey observing the central area of the Cosmic Evolution Survey \citep[COSMOS; ][]{Scoville2007} field which is one of the most popular deep multi-wavelength survey fields. It covers a contiguous 0.54 deg$^2$ region in four Near Infrared Camera (NIRCam) filters (F115W, F150W, F277W, and F444W), reaching a $5\sigma$ point source depth of $27.5 - 28.2$ magnitudes.  In parallel, a 0.19 deg$^2$ area of Mid-Infrared Instrument (MIRI) imaging with the F770W filter is covered. For this work, we make use of the 0.28 deg$^2$ \textit{JWST}/NIRCam F150W images reduced by \citet{Zhuang2024}. They used version 1.10.2 of the \texttt{JWST}\footnote{\url{https://jwst-pipeline.readthedocs.io/en/latest/}} pipeline with the Calibration Reference Data System (CRDS) version of 11.17.0. to reduce the uncalibrated NIRCam raw data retrieved from MAST\footnote{\url{https://archive.stsci.edu/doi/resolve/resolve.html?doi=10.17909/6btv-br09}}. The steps followed in the data reduction process are briefly summarised below:
\begin{enumerate}
\item Individual exposures of raw data were reduced using the Stage 1 pipeline Detector1Pipeline, in which they used some custom parameters to better flag large cosmic ray events and snowballs. 
\item Fully calibrated individual exposures were obtained from running the Stage 2 pipeline Image2Pipeline, which performed wcs assignments, flat-fielding and photometry calibration. 
\item A two-dimensional background was subtracted, after masking bad pixels and sources using SextractorBackground in the \verb|photutils| package \citep{Bradley2024_10967176}. 
\item Wisps (artefacts caused by scattered light in the mirror) and claws (features caused by scattered light coming from extremely bright stars) were subtracted. 
\item Finally, single mosaics for each filter were produced using the Stage 3 Image3Pipeline, by combining all calibrated images. 
\end{enumerate}
For each NIRCam mosaic, \citet{Zhuang2024} constructed three different PSF models. For this work, we used the global PSF models, which are produced using all of the point-like sources across the entire field-of-view (FoV) of each dither-combined mosaic. The median PSF full width at half maximum (FWHM) in the COSMOS-Web NIRCam mosaics is 61.1 mas for the F150W filter. The PSF in the NIRCam imaging has a fractional root mean square (RMS) temporal variation of PSF FWHM of $\sim2.4\%$ for F150W, which is dominated by short timescale fluctuation. In comparison, the spatial variation of PSF FWHM, dominated by random variations, is much larger, at a level of $\geq5\%$ for short-wavelength filters including F150W. For a full description of their data reduction method, we refer the reader to Sect. 2.1 of \citet{Zhuang2024}. The reduced NIRCam F150W imaging data as well as the PSF models used in this work are available at \url{https://ariel.astro.illinois.edu/cosmos_web/}.
In Fig.\,\ref{fig.psf} we illustrate the variations of the adopted global PSF models. We first stacked all 80 PSF models and then calculated the mean and the standard deviation pixel by pixel (displayed in the left and central panels of Fig.\,\ref{fig.psf}, respectively). The right panel of Fig.\,\ref{fig.psf} shows the relative dispersion (or coefficient of variation), which is the standard deviation divided by the mean (pixel by pixel).

We use the COSMOS2020 \citep{Weaver2022} photometric catalogue to construct our sample of real \textit{JWST} galaxies, within the COSMOS-Web area. For that, we use the Farmer's version for the photometric catalogue. We use spectroscopic redshifts when available. Otherwise, we adopt the photometric redshifts (photo-$z$) listed in the COSMOS2020 catalogue, computed using the LePhare code \citep{Arnouts2002, Ilbert2006}. The photo-$z$ precision, given by the normalised median absolute deviation, is around $0.01\times(1+z)$ at $i < 24.0$ mag, and $0.03\times(1+z)$ at $24.0 < i < 27.0$ mag. Stellar masses ($M_*$) are derived by La Marca et al. 2025 (in prep.), using the spectral energy distribution fitting tool CIGALE \citep{Burgarella2005, Noll2009, Boquien2019}, including AGN models. These stellar masses are consistent with the stellar masses presented in the COSMOS2020 photometric catalogue obtained by LePhare. We calculate the median bias, $b$, and the median absolute deviation (MAD) between the two stellar mass estimates as follows:
\begin{equation}
    b = \text{median}(\Delta M_*),
\end{equation}
\begin{equation}
    MAD = 1.48\times \text{median} (|\Delta M_* - \text{median}(\Delta M_*)|)\, ,
\end{equation}
where $\Delta M_* = log_{10}M_{*,LePhare} - log_{10}M_{*,CIGALE}$. We find $b=-0.09$ and $MAD=0.013$. We construct our galaxy sample in the redshift range $0.5<z<3$, avoiding stars or areas affected by bright stars. Finally, we select a stellar mass-complete sample by including galaxies with $M_{*}$ greater than either $10^9\,M_{\odot}$ or the $K_S$-based completeness limit from \cite{Weaver2022} for the COSMOS2020 catalog:
\begin{align}
M_{\mathrm{lim}}(z) &= -3.55 \times 10^8 (1+z) \nonumber \\
&\quad + 2.70 \times 10^8 (1+z)^2,
\end{align}
whichever is higher. The minimum stellar mass limit of $10^9\,M_{\odot}$ is imposed to match the limit used for the simulated galaxy sample. In total, we include 25\,596 galaxies in our stellar mass complete sample over the 0.28 deg$^2$ analysed in this study.

\subsection{\label{subsc:data.agn_selection}AGN selections}

We use the following two common AGN selection techniques to identify AGN in our \textit{JWST} sample and compare with our DL-based methodology in Sect. \ref{sec.results.jwst},

\begin{enumerate}

    \item X-ray AGN selection: For this selection, we use the X-ray photometry data from the {\it Chandra} COSMOS Legacy survey \citep{Civano2016, Marchesi2016}. From the catalogue of optical and infrared counterparts provided by \cite{Marchesi2016}, we select sources with a final counterpart identification flag of 1 (secure) or 10 (ambiguous). We then cross-matched this catalogue using the optical coordinates with our selected sources in COSMOS2020. We selected secure X-ray AGN requiring \textsc{det\_ml} (the maximum likelihood detection) >10.8 in the hard or soft band. This selection yields a total of 223 X-ray AGN within our final sample.
    
    \item MIR AGN selection: For this selection, we use the {\it Spitzer}/MIPS $24\mu m$ data, provided by the COSMOS-Spitzer programme \citep{Sanders2007}, and the MIR data from the four channels of the {\it Spitzer}/IRAC Cosmic Dawn Survey (Euclid Collaboration et al. 2022). We follow \citet{chang_infrared_2017} to identify AGN by their MIR emission. We limited the selection to those galaxies with \emph{Spitzer/}MIPS 24 $\mu$m flux $F_{24\mu m}>20\, \mu$Jy, which is the $1\sigma$ total noise (instrument and confusion noise). Additionally, we required the signal-to-noise ratio ($S/N>5$) in each IRAC channel. Then, we applied the colour$-$colour criteria in \citet{chang_infrared_2017} to select AGN:
\begin{align}
    y &< 2.22\times x +1.01, \\
    y &< 8.67 \times x -0.28, \\
    y &> -0.33 \times x +0.17, \\
    y &> 0.31\times x -0.06,
\end{align}
where $x=m_{3.6\mu m} - m_{5.8 \mu m}$ and $y=m_{4.5\mu m} - m_{8 \mu m}$. Magnitudes are expressed as AB magnitude. Applying these criteria results in 680 MIR-selected AGN in our final sample.
\end{enumerate}

\subsection{Mock \textit{JWST} images}\label{subsec.data.mock}

 \begin{figure*}
     \centering
     \includegraphics[width=0.84\textwidth]{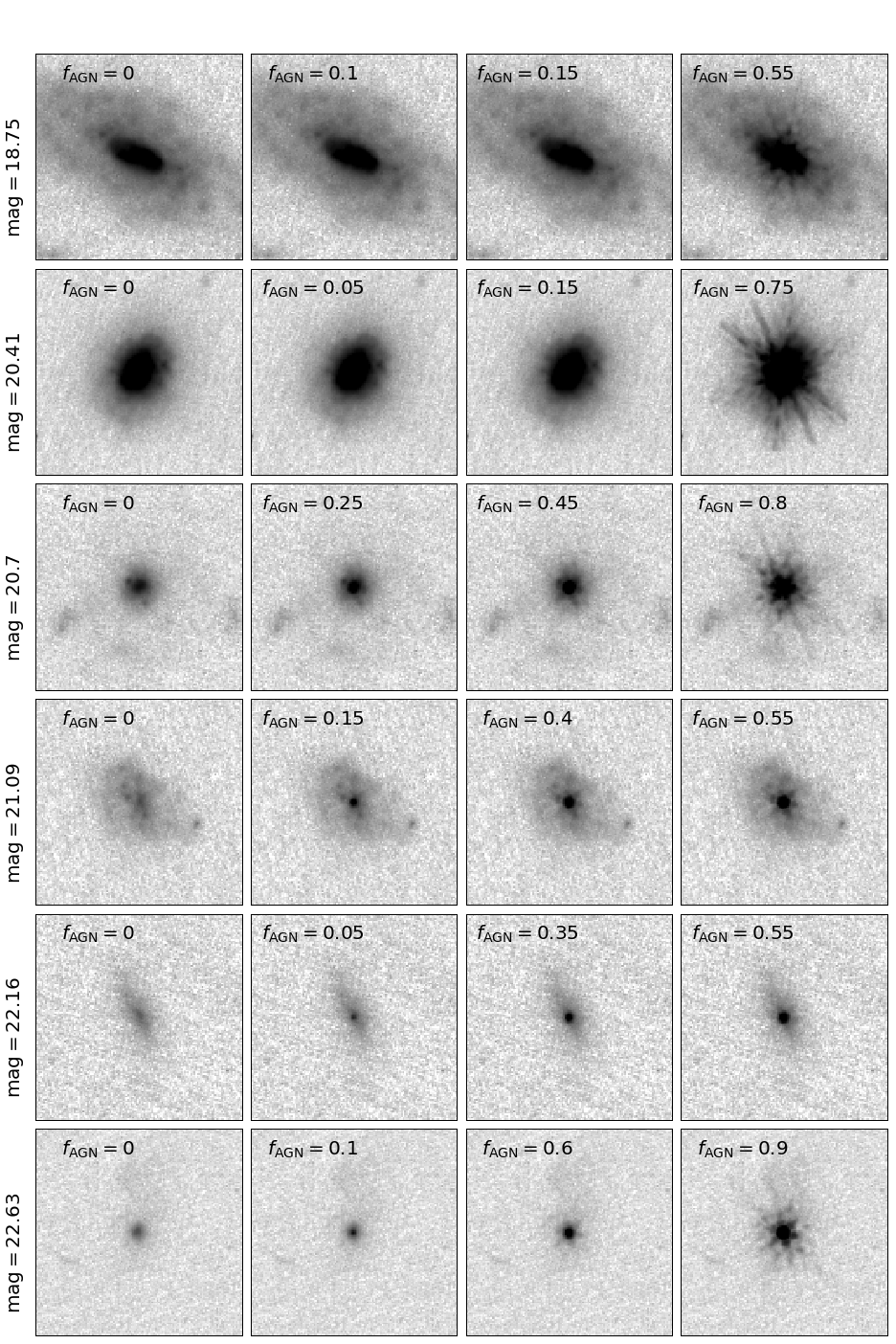}
     \caption{Example mock \textit{JWST}/NIRCam F150W images with varying levels of AGN contribution. The images have been generated to mimic \textit{JWST} observations, and include realistic \textit{JWST} noise and background. Each row corresponds to a different galaxy with no AGN contribution in the left panel and increasing AGN contributions in the rest of the panels. We show four example galaxies with different magnitudes, from the brightest (top) to the faintest (bottom). Images are 3.84\arcsec across and are displayed with an inverse arcsinh scaling. }
     \label{fig:injected_AGN}
 \end{figure*}

The IllustrisTNG project \citep{Nelson2019, Pillepich2018, Springel2018, Nelson2018, Naiman2018, Marinacci2018} is a series of cosmological hydrodynamical simulations of galaxy formation and evolution, with three different runs that differ in volume and resolution. These runs are TNG50, TNG100, and TNG300, with comoving length sizes of 50, 100, and 300 Mpc $h^{-1}$, respectively. The initial conditions for all runs are drawn from the Plank results \citep{Plank2016}. For this work, we used TNG100 (better resolution than TNG300 but still with a big enough volume to contain a sufficiently large number of galaxies), which contains $1820^3$ dark matter (DM) particles with a mass resolution of $M_{\mathrm{DM,\ res}} = 7.5 \times10^6 M_{\odot}$. In comparison, the baryonic particle resolution of TNG100 is $M_{\mathrm{baryon\,res}}=1.4\times10^6$ M$_{\odot}$. We refer the reader to \cite{Pillepich2018} for more details on IllustrisTNG. 

In this work, we selected galaxies from fourteen simulation snapshots (numbers 67, 64, 61, 58, 55, 52, 49, 46, 43, 40, 37, 33, 29 and 25), which correspond to redshifts from $z=0.5$ out to $3$. The time step between each snapshot is roughly $\sim$480\ \text{Myr} over this redshift interval. We selected all galaxies from these snapshots with stellar mass $M_{*}>10^9M_{\odot}$ to ensure that most galaxies have a sufficient number of stellar particles (hence reasonably well resolved),  with the lowest mass galaxies in TNG100 ($M_{*}=10^9M_{\odot}$) consisting of $714$ particles. We randomly choose a total of $\sim12000$ galaxies from this selection in mass and redshift to create our training sample. Every galaxy in IllustrisTNG has a complete merger history available from applying the SUBLINK algorithm on baryon-based structures \citep{Rodriguez2015}. We used these merger trees to identify major mergers and non-merger galaxies. Specifically, major mergers are defined as galaxies with stellar mass ratios > 1:4 and that will either have a merger event in the following 0.8 Gyr (pre-mergers) or had a merger event in the last 0.3  Gyr (post-mergers). Consequently, galaxies that do not satisfy these conditions are considered as non-mergers.

For each galaxy, we generated a synthetic \textit{JWST}/NIRCam F150W observation from the simulations with the following steps: 
\begin{itemize}
\item First, we created a smoothed 2D projected map \citep{Rodriguez-Gomez2019, Martin2022}. For each stellar particle in the simulation, we assigned a SED based on its mass, age, and metallicity, using the \citep{Bruzual2003} stellar population synthesis models, with a \cite{Chabrier2003} initial mass function. These SEDs were integrated through the \textit{JWST}/NIRCam F150W filter transmission curve to obtain the flux contribution of each particle in that band. The spatial distribution of flux was smoothed using an adaptive kernel to create the 2D map. This approach does not include full radiative transfer calculations, and therefore dust attenuation and emission lines are not modeled. The images were produced with the same pixel resolution as the real \textit{JWST} observations (0.03 \arcsec/pixel), and have a physical size of $50\times50$ kpc. 
\item Second, each image was convolved with a randomly chosen global \textit{JWST} F150W PSF model (out of 80 in total, as derived by \citealt{Zhuang2024}). 
This step ensures that our training sample contains the full information on the (spatial and temporal) variation of the PSF. 
\item Third, Poisson noise was added to each image, to account for the statistical variation of a source's photon emissions over time. 
\item Lastly, each image was injected into cutouts of real \textit{JWST} F150W sky cutouts. This step ensures that fully realistic background and noise are included in our training data.
\end{itemize}
In the final step, we followed the same approach as in \citet{Margalef2024}. To obtain cutouts of the real \textit{JWST} sky, we first generated a catalogue of sources that we wanted to avoid within the central region of the cutouts. Starting from the `Farmer' version of the COSMOS2020 catalogue \citep{Weaver2022}, we set the flag \textsc{flag\_combined} equal to 0 to select areas that are not affected by bright stars or large artefacts, as recommended by the COSMOS team. We then made use of the star/galaxy separation provided in the Le Phare photo-$z$, selecting sources with the star/galaxy flag lp\_type equal to 0 (galaxy) or 2 (X-ray source). Finally, we restricted our selection only to $z<3$ sources in the COSMOS-Web field. 
The final catalogue contains relatively bright sources, with average magnitudes being: HSC-$i=25.6$ mag, UVISTA-$H=24.7$ mag.
Then, we generated random sky coordinates such that there are no catalogued bright sources within a circular radius of $6.5\arcsec$. This radius corresponds to the estimated source density of the area from which we extracted the cutouts. These criteria ensure that there are no bright sources in the centre of the cutouts, where the synthetic galaxies will be injected, but still allow for faint background galaxies. The random coordinates were then used as the centres for the sky cutouts, in which to inject the simulated galaxies, after performing sanity checks to ensure that no artefacts, stellar spikes, or bad or saturated pixels were present. 

To have a uniform dataset to train our DL model, and to speed up computation, we cut all images to $128\times128$ pixels, without changing the pixel scale of 0.03\arcsec/pixel. In other words, the images are 3.84\arcsec across, corresponding to roughly 23 - 33 kpc in physical scale over the redshift range of our sample (i.e. $0.5<z<3$).

\section{Methods}\label{sect:Methods}

\begin{figure*}
    \centering
    \includegraphics[width=\linewidth]{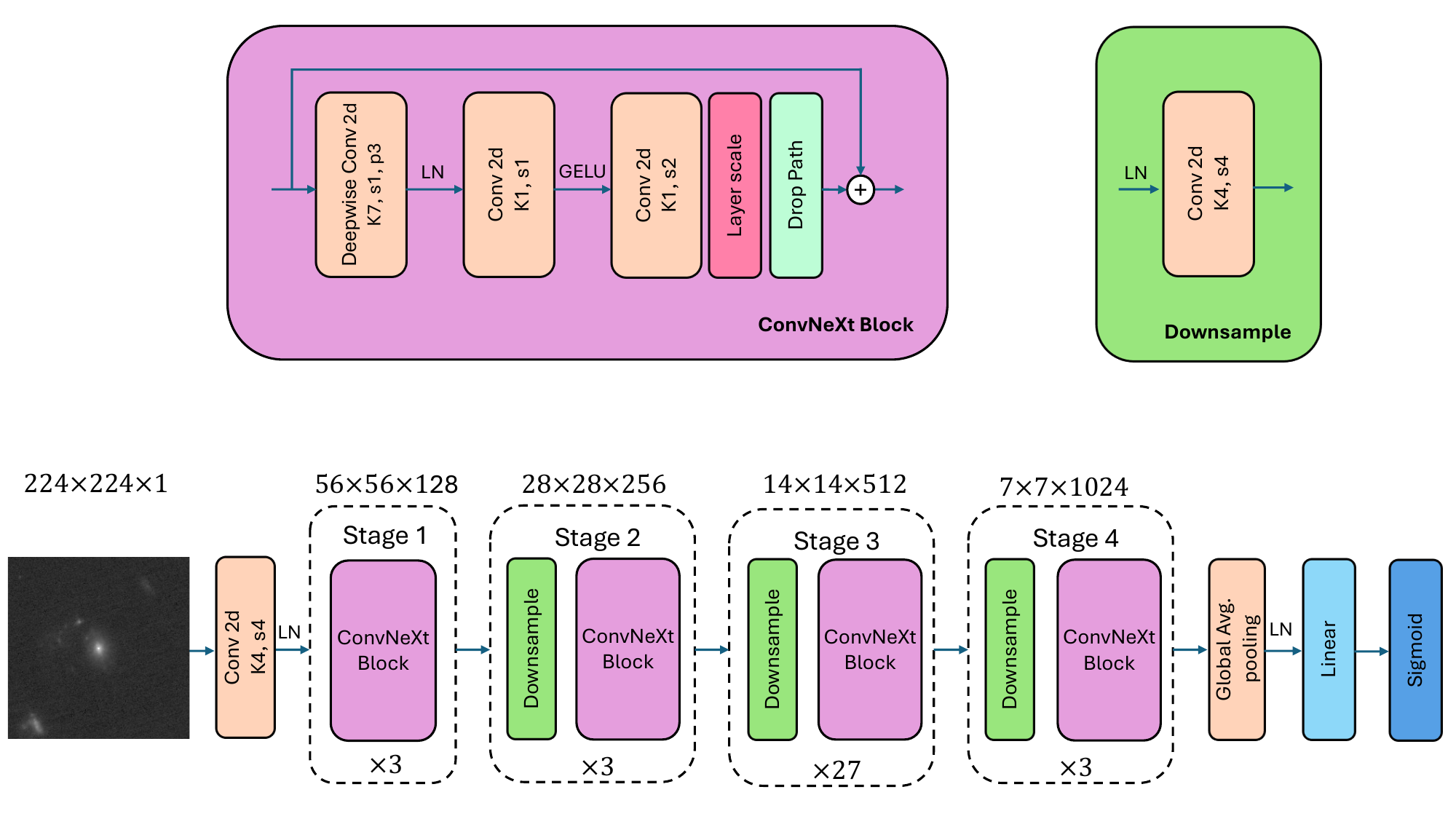}
    \caption{Architecture of ConvNeXt-Base network (bottom) with a four-stage feature hierarchy, which allows us to extract features on different scales. On top of each stage, we show the dimension of the feature maps, with the width and height decreasing as the network deepens while the filter size increases. The top left diagram shows the internal structure of ConvNeXt Block. The top right diagram shows the internal structure of Downsample. The LN and GELU represent a layer normalisation and a Gaussian error linear unit activation function, respectively.}
    \label{fig.network}
\end{figure*}

In this section, we first describe the construction of the mock \textit{JWST}/NIRCAM host galaxy images with different injected levels of AGN contribution. Then we introduce the two methods (our method based on DL and GALFIT based on 2D surface brightness fitting) used to estimate the AGN contribution fractions.

\subsection{Mock AGN injection}\label{sect:mock.agn}

To simulate images of galaxies with AGN, we injected a central point source into the host galaxy image. The observed \textit{JWST} PSF models as described in Section \ref{subsec.data.jwst} were used as the central point source.  In order to create different AGN contribution fractions, the relative brightness of the PSF was adjusted before injecting it into the mock \textit{JWST} images described in Section \ref{subsec.data.mock}. 

The AGN contribution fractions were chosen to range from 0.0 to 0.95 in increments of 0.05, for a total of 20 different levels. For any given AGN contribution fraction $f_{AGN}$, the injected image was created as follows. 
First, the PSF image was transformed into a pixel scale of 0.03\arcsec/pix, same as the simulated images. 
Then, the flux of the PSF image was measured within a 2\arcsec aperture using the \verb|aperture_photometry| function of the \verb|photutils| package \citep{Bradley2024_10967176}. This aperture size corresponds to sizes between $13$ kpc and $18$ kpc in our redshift range. The PSF image was normalised using this flux, so it could be easily adjusted later. The flux of the host galaxy image, $F_{host}$, was measured in the same way. The AGN contribution fraction (i.e. contribution to the total light) is then defined as 
\begin{equation}
    \label{eq:fAGN}
    f_{AGN} = \frac{F_{AGN}}{F_{host} + F_{AGN}},
\end{equation}
 where $F_{AGN}$ is the flux of the AGN. Afterwards, the normalised PSF image was multiplied by $F_{AGN}$, which can be derived from Eq. \ref{eq:fAGN}:
 \begin{equation}
     F_{AGN} = \frac{f_{AGN}}{1-f_{AGN}} F_{host}.
 \end{equation}
Finally, the scaled PSF image was added to the host galaxy image. 

For each galaxy, five different images were created with five different AGN contribution fractions, chosen randomly from the 20 possible discrete AGN contribution fractions. Example images of these simulated galaxies (from bright to faint) with varying levels of AGN contribution can be seen in Fig.\,\ref{fig:injected_AGN}. Visually, it is clear that it can be very difficult to discern the presence of the host galaxy when the AGN contribution fraction is large, particularly for fainter and higher-redshift objects. Later, we will examine how well we can recover $f_{AGN}$ as a function of the intrinsic AGN contribution fraction, redshift, galaxy size, S/N, etc, using a set of common performance metrics.

 \begin{figure}
     \centering
     \includegraphics[width=0.5\textwidth]{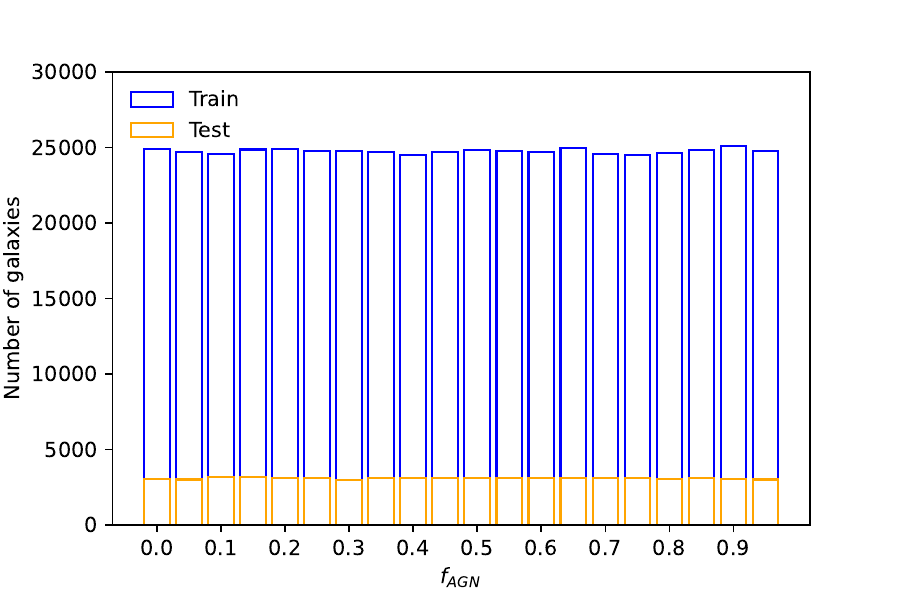}
     \caption{Distributions of the injected AGN contribution fraction (as defined in Eq. \ref{eq:fAGN}) in the training (blue) and test datasets (orange). The distributions are mostly uniform for both the training and test datasets.}
     \label{fig.train_test_data}
 \end{figure}

\subsection{Deep learning CNN}\label{sect:CNN}

Zoobot \citep{Walmsley2023} is a Python package used to measure detailed morphologies of galaxies (such as spiral arms, bars and bulges) using DL based on the idea of successive layers of learned representations. Zoobot includes convolutional neural networks \citep[CNNs,][]{Fukushima1988, LeCun2015} and vision transformer models \citep{Dosovitskiy2021, Dehghani2023}. These models are pre-trained on many millions of labelled galaxies, derived from the visual classifications of the Galaxy Zoo project\citep{Lintott2008} on real images of galaxies selected from surveys such as the Sloan Digital Sky Survey (SDSS), Hyper Suprime-Cam (HSC) and Hubble \citep{Willett2013, Willett2017, Simmons2017, Walmsley2022a, Walmsley2022b}. The models are designed to be easily adaptable to new tasks (classification or regression tasks) and galaxy surveys with a minimal amount of new labelled data.

Some of the available models in Zoobot belong to the family of ConvNeXts \citep{Liu2022}. 
They are pure convolutional models constructed by optimising the ResNet \citep{He2015} architecture to bear resemblance with vision transformers \citep{Vaswani2017, Liu2021}, in which the design choices such as the use of the Gaussian Error Linear Unit (GELU) activation function or inverted bottleneck CNN blocks (which are a specialised type of residual block more computationally efficient than normal residual blocks) are proven to improve the performance of a purely CNN model and can compete with transformer models in terms of accuracy and scalability. 

For this work, we chose a ConvNeXt-Base architecture, which was pre-trained on the Galaxy Zoo dataset of over 820k images and 100 million volunteer votes on morphological questions. In order to perform a regression task (as is the case in this study), we added a linear head with a sigmoid function (to restrict the output to be between 0 and 1) and used a mean square error loss function to train the network. A diagram of the ConvNeXt-Base network can be seen in Fig.\,\ref{fig.network}. For this work, we retrained the last two blocks of the network and the linear head, while the rest of the network parameters were frozen to keep the optimal values found for the pre-trained data from Galaxy Zoo. Our model was trained on a v100 GPU and took 72 hours to complete. Once the model is trained, it takes Zoobot $6\times10^{-3}$ seconds to predict one galaxy.

We split our sample of mock \textit{JWST} images in training and test sets, with a 90/10 split. The distribution of the injected AGN contribution fraction in the training and test datasets can be seen in Fig.\,\ref{fig.train_test_data}. The training dataset was used for training Zoobot. A validation set of 10\% of the training data was used to monitor the performance during training. The test set was used to determine the performance of the final model on data that had never been seen before by the algorithms. To ensure the test set cannot be learned by simply interpolating from the training set, the galaxies were split into the train, validation and test sets in a way that the five iterations of the same galaxy (with different injected AGN contribution fractions) were only used in one of the splits. Furthermore, we randomly selected a subset of $4800$ galaxies from the test set on which to directly compare the results from the DL methodology and surface brightness fitting with GALFIT. This subset was constructed in a way that for each snapshot we randomly selected 400 galaxies with a uniform distribution of $f_{AGN}$.

\subsection{GALFIT}\label{sect:Galfit}

\begin{figure}
    \centering
    \includegraphics[width=\linewidth]{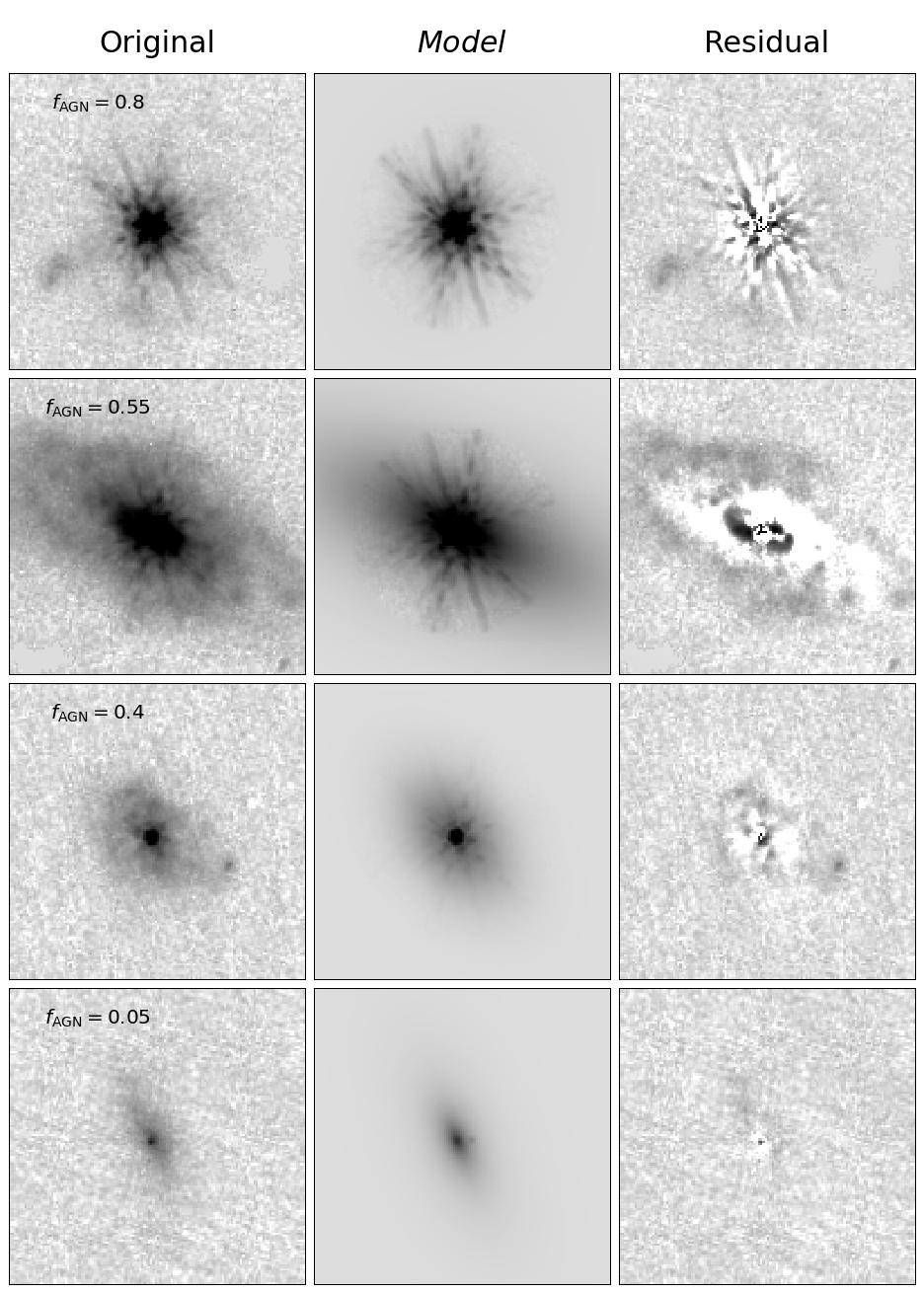}
    \caption{S\'ersic + PSF decomposition of four example galaxies (with AGN contributon fraction varying from high to low from top to bottom) on which we performed GALFIT. Images of the original galaxy, the model (S\'ersic + PSF), and the residual (original$-$model) are shown from left to right.  Images are 3.84\arcsec across, and are displayed with an inverse arcsinh scaling. }
    \label{fig.example_galfit}
\end{figure}

GALFIT \citep{Peng2002} is a popular two-dimensional fitting code used to model the surface brightness of an object with pre-defined analytic functions. GALFIT allows the user to fit any number of components and different light profiles (e.g. S\'ersic, exponential disk, PSF, etc.). The best-fit model is obtained by $\chi^2$-minimisation using a Levenberg-Marquardt algorithm. A S\'ersic profile generally describes well the light distribution of spheroidal or disk galaxies (even though it will not be able to represent well disturbed or irregular galaxies). It has the following functional form,

\begin{equation}
    \sum (R) = {\sum}_e exp \left\{ -\kappa_n  \left[ \left( \frac{R}{R_e} \right)^{1/n} -1 \right] \right\}, 
\end{equation}

\noindent where $R_e$ is the effective radius, 
such that half of the total flux is contained within $R_e$, $\sum_e$ is the surface brightness at $R_e$, $n$ is the S\'ersic index (it determines the shape of the light profile, and $n=1$ represents a disk while $n=4$ represents a spheroid), $\kappa_n$ is a positive parameter that depends on $n$. On the other hand, the PSF profile can be used to describe a central point source such as the AGN. We can fit the surface brightness of each galaxy with a combination of S\'ersic and PSF profiles in order to derive the contribution from a possible AGN component at the centre of the galaxy.

We ran GALFIT on the subset of 4800 galaxies from the test set. We first ran Sextractor \citep{Bertin1996} on all the test images to determine the central positions in pixels, x and y, the total magnitude of the galaxy, axis ratio (q) and effective radius ($R_e$). Sextractor also produces a segmentation map in which all galaxies are identified and can be used to either mask neighbouring galaxies or fit them simultaneously with the central galaxy. In order to perform $\chi^2$-minimisation, GALFIT requires a sigma map. This sigma map was constructed by adding the Poisson noise contribution from the simulated galaxy (after injecting the AGN contribution) to the error map provided by the \textit{JWST} data. 

We ran GALFIT with a combination of a single S\'ersic profile and a PSF model for the main central galaxy in each image. Neighbouring galaxies were masked, unless their light overlaps with the central galaxy, in which case they were fitted simultaneously with a S\'ersic profile. The initial estimates of the model parameters can have an impact on whether GALFIT finds a good fit or not. That is why we ran GALFIT with different initial parameters of S\'ersic index and magnitudes. We chose the best model to be the one with the lowest reduced $\chi^2$. For the S\'ersic index of the main galaxy, we chose as initial values $n=1,2,4$, alternatively. For the magnitudes of the S\'ersic and PSF models, we used three different combinations of the initial parameters. In the first scenario, both models were set to be equal to a magnitude that corresponds to half of the total flux obtained from Sextractor. In the other two scenarios, the magnitude of the PSF (S\'ersic) model was set to be 80\% of the total flux while the magnitude of the S\'ersic (PSF) model was set to be 20\%. For the rest of the parameters (position x and y, axis ratio, position angle and effective radius), we used as initial values those obtained by Sextractor. In total, this resulted in 9 different model configurations (3 S\'ersic index values $\times$ 3 magnitude combinations), from which we selected the one with the lowest reduced $\chi^2$ as the final best fit. For the neighbouring galaxies that were fitted simultaneously, we also used the Sextractor parameters plus a S\'ersic index of $n=2$. On average, GALFIT took 15 seconds per galaxy to complete the fitting procedure (2500 times slower than the prediction time from Zoobot), which resulted in a total of 90 hours to fill all 9 model configurations (resulting from the different combinations of initial S\'ersic indices and magnitudes) to the 2400 galaxies in our test subset.

In some cases, GALFIT did not converge and did not produce a fit at all. In other cases, even if GALFIT produced an output, it was clearly not a good fit for the surface brightness light. We only selected galaxies for which GALFIT produced a good fit, that is, it has a reduced $\chi^2 < 5$ and no non-physical parameters (for example effective radius smaller than 0.5 pixels, or larger than the size of the image stamp, $q < 0.1$, and $n < 0.5$ or $n > 10$). In Fig.\,\ref{fig.example_galfit}, we present good GALFIT fits of example galaxies containing varying amounts of injected AGN contribution fractions (from insignificant to dominant AGN contributions), with the original image, the model image (S\'ersic + PSF) and the residual image (original - model) shown in different columns. After the best fit had been found, we created images of each model separately (S\'ersic and PSF) using the parameters from the best fit. We then calculated the flux within an aperture of 2\arcsec in the S\'ersic model, the PSF model and the original galaxy image. 

Finally, we calculated the AGN contribution fraction derived from running GALFIT in two slightly different ways. In the first method, we set the derived AGN contribution fraction to be equal to the ratio of the aperture flux from the PSF component to the aperture flux of the original galaxy image (which corresponds to the total flux of the galaxy within a 2\arcsec aperture). In the second method, we used as total flux (within a 2\arcsec  aperture) the sum of the aperture fluxes of the PSF and the S\'ersic model. We adopted these two different approaches to calculating the total flux of the galaxy in order to understand if there is any systematic bias in the S\'ersic model, which could also impact the flux of the PSF model.

\section{Results}\label{sect:Results}

In this section, we first present results from our DL-based method Zoobot. We analyse in detail how key performance metrics such as the root mean square error (RMSE), relative absolute error (RAE) and outlier fraction vary as a function of the injected AGN contribution fraction, redshift, S/N, and galaxy size. Then, we compare these results with the performance obtained from running GALFIT on the same test set. Finally, we apply our DL-based method to the real \textit{JWST} images of the stellar mass complete galaxy sample and compare AGN identified using our method with AGN selected in the X-ray and MIR.

\subsection{Zoobot model performance}\label{subsec.Zoobot}

\begin{figure}
    \centering
    \includegraphics[width=0.48
\textwidth]{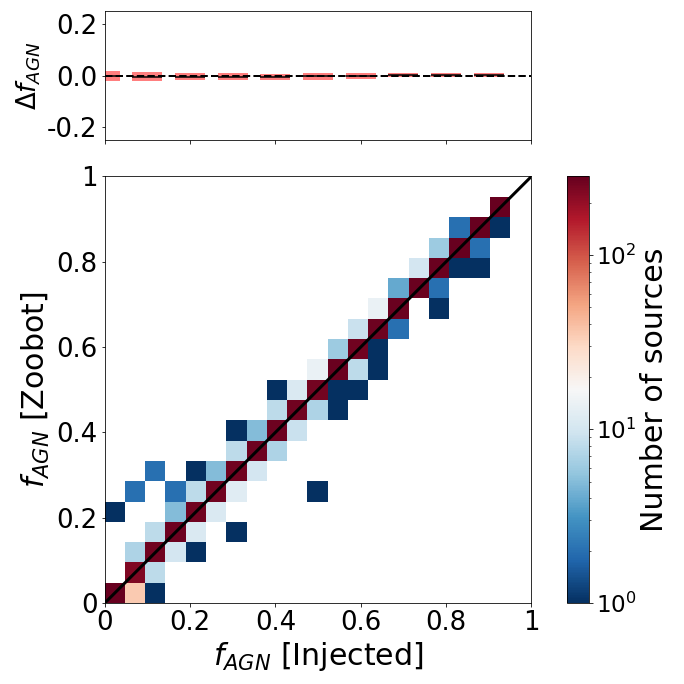}
    \caption{Comparison between the real injected AGN contribution fraction and the AGN contribution fraction obtained from the Zoobot model on the subset of 4800 galaxies across the whole redshift range ($0.5<z<3$). The comparison shows a mean difference between the two quantities ($\Delta f_{AGN} = f_{AGN}$ [Injected] $- f_{AGN}$ [Zoobot]) of $-0.0018$ and an overall RMSE = 0.013. The solid diagonal line is the 1:1 line. The top plot shows the mean difference and its dispersion as a function of the injected AGN contribution fraction. The colour bar indicates the number of sources in each bin.}
    \label{fig_pred_zoobot}
\end{figure}

\begin{table*}[]
\caption{Performance statistics for Zoobot and GALFIT.}
    \centering
    \begin{tabular}{|l|c|c|c|c|c|c}
    \hline
                                                       &  RMSE             & RAE & $\langle{\Delta f_{AGN}}\rangle$ & Outlier (20\%)    & Outlier (30\%)    \\
               \hline
        Zoobot                                         & $0.013 \pm 0.005$ & $0.076 \pm 0.005$ & $-0.0018 \pm 0.0002$ & $0.065 \pm 0.004$ & $0.061 \pm 0.003$ \\
        Zoobot$^{M}$                                   & $0.014 \pm 0.006$ & $0.102 \pm 0.009$ & $-0.0017 \pm 0.0005$ & $0.072 \pm 0.010$ & $0.065 \pm 0.010$ \\      
        Zoobot$^{NM}$                                  & $0.013 \pm 0.006$ & $0.073 \pm 0.005$ & $-0.0018 \pm 0.0002$ & $0.065 \pm 0.004$ & $0.061 \pm 0.004$ \\  
        GALFIT ($f_{host}$ from original image)        & $0.12 \pm 0.07$ & $0.19  \pm 0.01$  & $-0.018  \pm 0.002$  &  $0.19 \pm 0.01$  & $0.13 \pm 0.01$ \\   
        GALFIT$^{M}$ ($f_{host}$ from original image)  & $0.13 \pm 0.07$ & $0.27  \pm 0.05$  & $-0.024  \pm 0.004$  &  $0.25 \pm 0.03$  & $0.18 \pm 0.02$ \\
        GALFIT$^{NM}$ ($f_{host}$ from original image) & $0.12 \pm 0.07$ & $0.17  \pm 0.01$  & $-0.016  \pm 0.002$  &  $0.16 \pm 0.01$  & $0.11 \pm 0.01$ \\
        GALFIT  ($f_{host}$ from model)                & $0.17 \pm 0.08$ & $0.51  \pm 0.02$  & $-0.075  \pm 0.003$  &  $0.46 \pm 0.01$  & $0.35  \pm 0.01$  \\
        GALFIT$^{M}$ ($f_{host}$ from model)           & $0.16 \pm 0.09$ & $0.48  \pm 0.07$  & $-0.087  \pm 0.007$  &  $0.54 \pm 0.04$  & $0.39  \pm 0.04$  \\
        GALFIT$^{NM}$ ($f_{host}$ from model)          & $0.17 \pm 0.08$ & $0.52  \pm 0.02$  & $-0.073  \pm 0.003$  &  $0.45 \pm 0.02$  & $0.34  \pm 0.01$  \\
        \hline
    \end{tabular}
    \tablefoot{We summarise the overall performance from Zoobot and GALFIT (for both ways of calculating the AGN fraction contribution), in terms of the RMSE, RAE, mean difference and outlier fractions (at different percentage levels, 20\% and 30\%). We show the results for the whole sample and mergers (M) and non-mergers (NM) separately.}
    \label{tab.results}
\end{table*}

Here we analyse the performance of the Zoobot model on the subset of 4800 galaxies from the test set. We first calculate the root mean square error (RMSE),
\begin{equation}
\textrm{RMSE} = \sqrt{\frac{1}{n}\Sigma^{n}_{i=1} (f^i_{AGN} [\textrm{injected}] - f^i_{AGN} [\textrm{predicted}])^2},
\end{equation}
which measures the average difference between the predicted values from a model (e.g. Zoobot or GALFIT) and the actual injected values. In Fig.\,\ref{fig_pred_zoobot} we show the predicted values of $f_{AGN}$, obtained from Zoobot, versus the real $f_{AGN}$ for all redshifts, which shows a very tight correlation across the whole dynamic range. Quantitatively, we found an overall value of $\text{RMSE}=0.013$ for the Zoobot model, demonstrating very good recovery (both in terms of accuracy and precision) of the injected real AGN contribution fractions. In addition, the top panel of Fig.\,\ref{fig_pred_zoobot} shows the mean and the dispersion of the difference between the real and predicted values ($\Delta f_{AGN} = f_{AGN}$ [injected] $- f_{AGN}$ [Zoobot]) as a function of the injected AGN fraction, with a value for the mean of $\langle{\Delta f_{AGN}}\rangle=-0.0018$, and dispersion of $\sigma(\Delta f_{AGN})=0.013$. This shows that there is very little bias from the Zoobot predictions to the actual values, and it is independent of $f_{AGN}$, as the difference between the real and predicted value over the whole range of $f_{AGN}$ is always below -0.0043. 
To check if the performance depends on galaxy structural properties, we further explore whether the S\'ersic index impacts how well Zoobot can predict the AGN contribution fraction (see Fig.\,\ref{fig.apendix1} in Appendix \ref{sec.appendix.sersic}) and find that for galaxies with $n<1$ the RMSE increases by 30\% (compared to the overal RMSE of 0.013), to RMSE$=0.017$ and that high S\'ersic indexes ($n>6$) do not lead to worse predictions (i.e., similar RMSE to the overal RMSE of $0.013$).

\begin{figure*}
    \centering
    \includegraphics[width=0.48\textwidth]{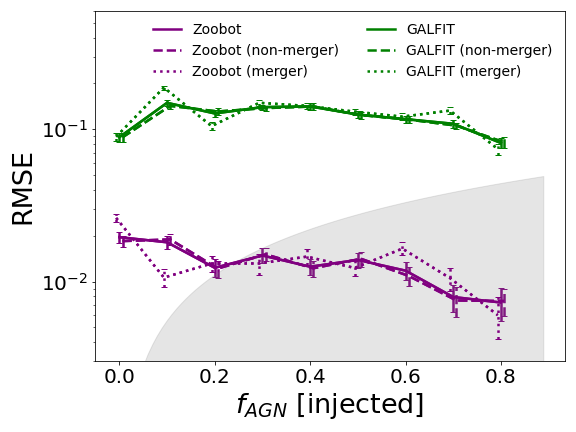}
    \includegraphics[width=0.48\textwidth]{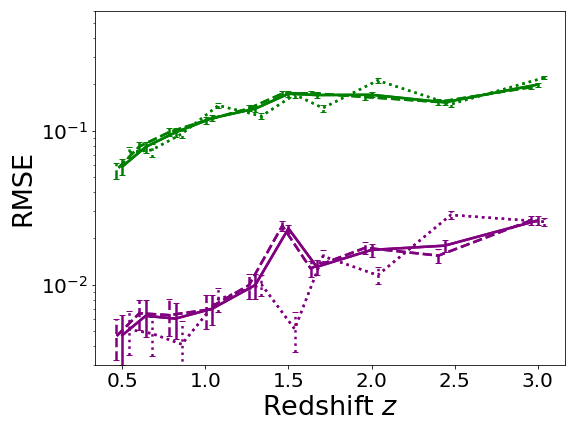}
    \includegraphics[width=0.48\textwidth]{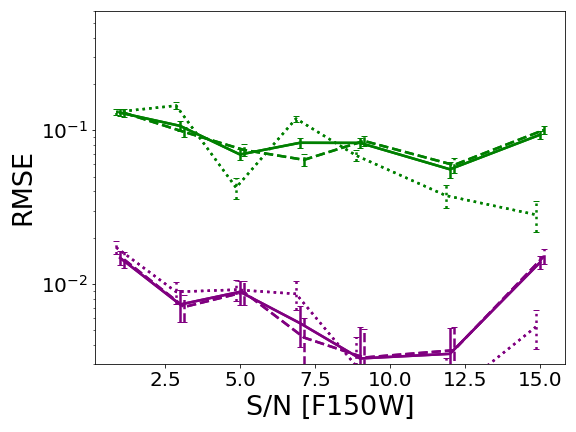}
    \includegraphics[width=0.48\textwidth]{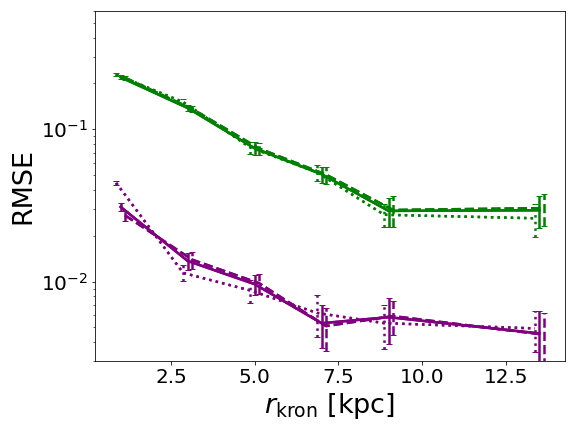}
    \caption{RMSE as a function of the injected AGN contribution fraction (top left), redshift (top right), S/N (bottom left) and $r_{kron}$ (bottom right). The purple lines correspond to the results from Zoobot and the green lines from GALFIT. The solid lines correspond to the whole sample while the dashed and dotted lines correspond to the mergers and non-merger galaxies, respectively. The error bars show the 95\% interval from bootstrapping the RMSE value. The performance of Zoobot is around a factor of 10 better than GALFIT.
    The shaded region in the top left panel represents the fractional variation (standard deviation divided by the mean) of the PSF, considering the spatial and temporal variations.
    }
    \label{rmse_z}
\end{figure*}

\begin{figure*}
    \centering
    \includegraphics[width=0.48\textwidth]{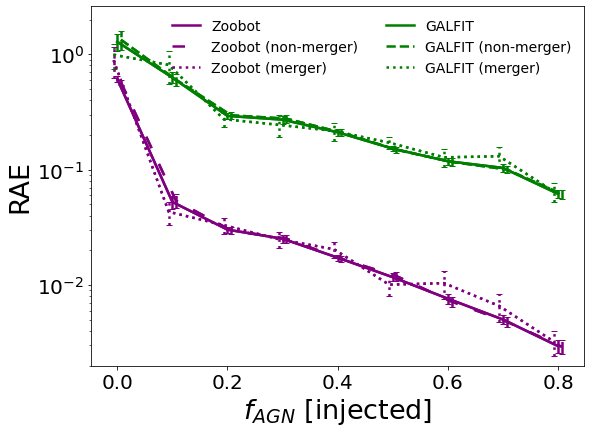}
    \includegraphics[width=0.48\textwidth]{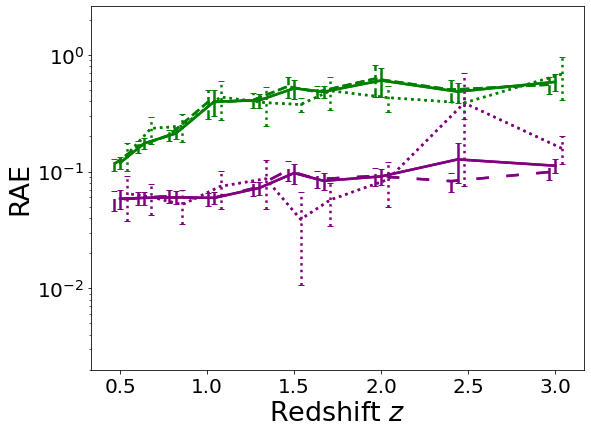}
    \includegraphics[width=0.48\textwidth]{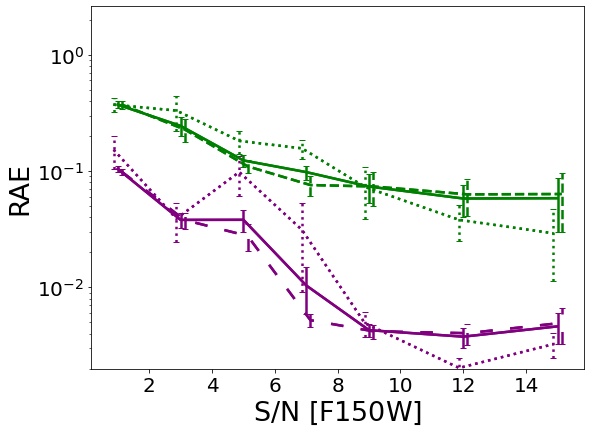}
    \includegraphics[width=0.48\textwidth]{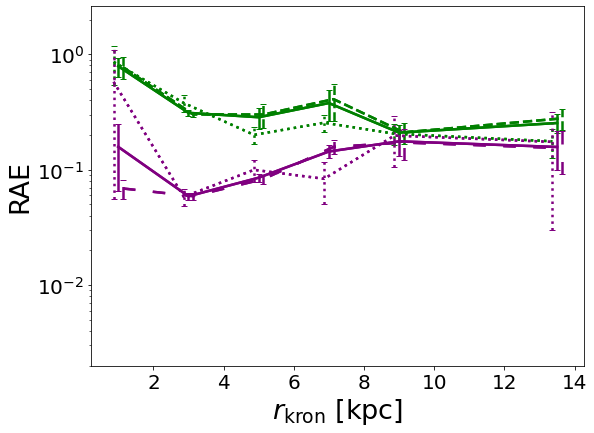}
    \caption{Average relative absolute error in bins of injected $f_{AGN}$ (top left panel), redshift (top right panel), S/N (bottom left panel) and $r_{kron}$ (bottom right), from the Zoobot predictions. The error bars correspond to the standard deviation of the data points in each bin. The purple lines correspond to the results from Zoobot and the green, from GALFIT. The solid lines correspond to the whole sample while the dashed and dotted lines correspond to the merger and non-merger galaxies, respectively. }
    \label{relative_error}
\end{figure*}

\begin{figure*}
    \centering
    \includegraphics[width=0.48\textwidth]{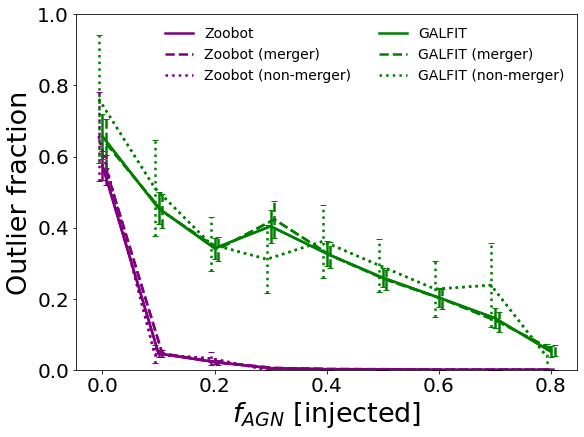}
    \includegraphics[width=0.48\textwidth]{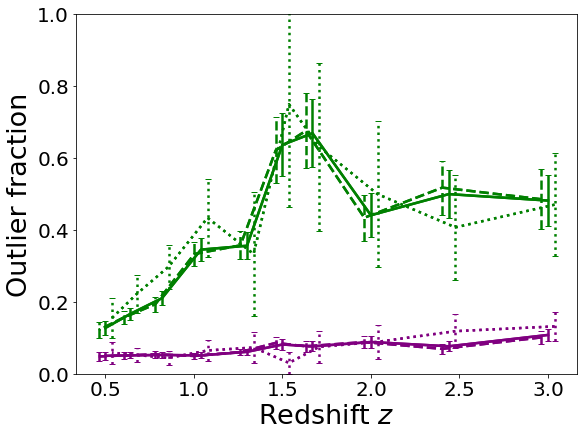}
    \includegraphics[width=0.48\textwidth]{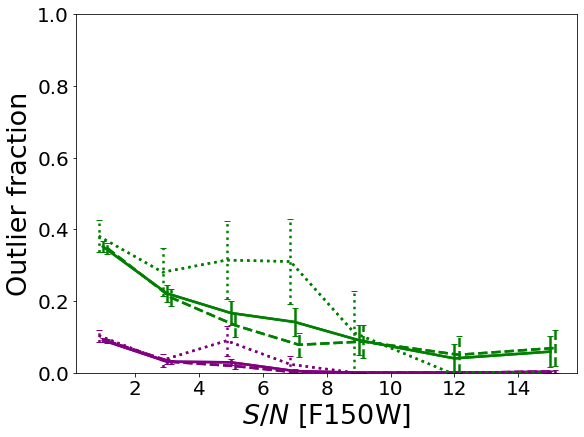}
    \includegraphics[width=0.48\textwidth]{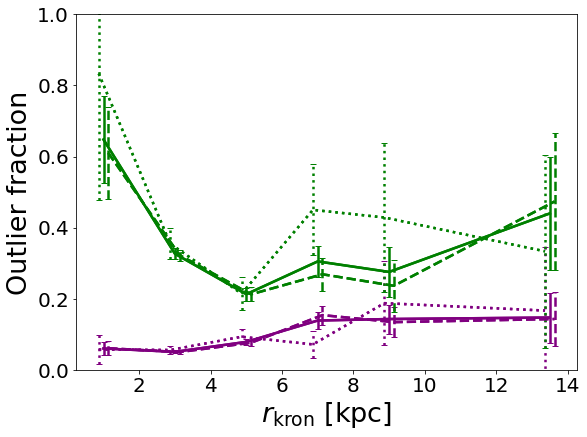}
    \caption{Outlier fraction (calculated as the fraction of galaxies with RAE $> 20\%$) as a function of the injected AGN contribution fraction (top left panel), redshift (top right), S/N (bottom left) and $r_{}$ (bottom right). The purple lines correspond to the results from Zoobot and the green, from GALFIT. The solid lines correspond to the whole sample while the dashed and dotted lines correspond to the merger and non-merger galaxies, respectively.}
    \label{outlier_fraction_20}
\end{figure*}

In Fig.\,\ref{rmse_z} we show the RMSE of the Zoobot predictions (purple lines) as a function of the injected AGN contribution fraction, redshift, S/N and kron radius ($r_{kron}$, calculated with SExtractor in
the F150W filter). The solid lines show the RMSE of the whole test subset. To separately investigate how the performance might change for highly disturbed or irregular galaxies, we also show galaxies classified as mergers (dashed lines) and galaxies classified as non-mergers (dotted lines), separately. In general, the RMSE is slightly lower for non-merging galaxies, but the difference with mergers is very small. 
The RMSE is larger for lower values of the injected AGN contribution fraction and reaches a maximum mean value of 0.020. 
At $f_{AGN}>0.2$, Zoobot achieves smaller errors than the coefficient of variation of the PSF, that is, smaller error than the error on the PSF resulting from the intrinsic temporal and spatial variations (shaded region of Fig.\,\ref{rmse_z}). In other words, the precision of our DL-based method is better than the intrinsic variations of the observed PSF. This is only possible because the DL method is trained on data which include the full range of observed PSFs.  
There is an increase in the RMSE with increasing redshift, with a maximum value of RMSE $=0.026$ at $z=3$, possibly because galaxies at higher redshifts are smaller and tend to have lower S/N (i.e. more dominated by noise). 
Indeed, we see that generally the RMSE increases with decreasing $r_{kron}$ and S/N. However, we also observe a small rise in RMSE at the highest S/N end, which is likely due to low number statistics in that bin; in such cases, even a single outlier can have a disproportionately large effect on the RMSE.

In Fig.\,\ref{relative_error} we show the relative absolute error (RAE), which is the ratio between the absolute error divided by the real value\footnote{The relative error is not well defined when the real AGN fraction is equal to zero. In order to calculate the relative error when $f_{AGN}$ [injected] $= 0$, we approximate it with $f_{AGN}$ [injected] $= 1e-3$.},
\begin{equation}
\textrm{RAE} = \frac{|f_{AGN}[\textrm{injected}] - f_{AGN} [\textrm{predicted}] |}{f_{AGN} [\textrm{injected}]}. 
\end{equation}
The purple lines (solid lines represent the whole test subset, dashed lines mergers and dotted lines non-merger) show the relative error from the Zoobot predictions as a function of the injected AGN contribution fraction, redshift, S/N and $r_{kron}$ (calculated by SExtractor in the F150W filter). We find that the relative error increases as $f_{AGN}$ decreases,  particularly for $f_{AGN} < 0.1$. While the RMSE only quantifies the absolute difference between real and predicted, the relative error does so in relation to the actual value. The relative error does not change much with redshift. But, similarly to the RMSE, it increases with decreasing S/N and decreasing $r_{kron}$, as expected. Overall, there is a small difference between mergers (dashed lines) and non-mergers (dotted lines), with non-mergers having slightly higher relative errors in some cases.

Finally, we explore the outlier fraction in the Zoobot predictions. Here we define outliers as the Zoobot predicted AGN contribution fractions with an RAE of more than 20\%, that is, 
\begin{equation}
\frac{|f_{AGN}[\textrm{predicted}] - f_{AGN}[\textrm{injected}]|}{f_{AGN}[\textrm{injected}]}>20\%. 
\end{equation}
In Fig.\,\ref{outlier_fraction_20}, we show the outlier fraction as a function of the injected AGN contribution fraction, redshift, S/N and $r_{kron}$ in purple lines. The overall outlier fraction for the whole sample is 6.5\%. At intermediate to high levels of AGN contribution, the outlier fraction is extremely low (close to zero). However, at low levels of injected AGN fractions, particularly for $f_{AGN}<0.1$, it increases very sharply with decreasing $f_{AGN}$. This behaviour is partly caused by our definition of outliers (i.e., it depends on whether we use a threshold on relative or absolute errors).
The outlier fraction remains more or less constant as a function of redshift and galaxy size. Concerning the S/N, there is only a small increase in the outlier fraction for $S/N < 3$. At $S/N>3$, the outlier fraction drops to almost zero. 
There is no significant difference in the outlier fractions between mergers (dashed lines) and non-mergers (dotted lines). 

In Table \ref{tab.results} we summarise the results obtained from Zoobot, in terms of the RMSE, RAE, mean difference and outlier fractions (at different percentage levels, 20\% and 30\%) for the whole test sample and for mergers and non-mergers separately.

\subsection{GALFIT performance}

We now analyse the results of the derived AGN contribution fraction after performing 2D light profile fitting on the same subset as Sec. \ref{subsec.Zoobot}. Each galaxy was fitted with a single S\'ersic and a PSF component, which describe the host galaxy and central source component, respectively. The AGN contribution fraction derived from this method was calculated by dividing the flux (within a circular 2\arcsec aperture) of the PSF model by the total flux of the galaxy (within the same aperture). As mentioned before, we can calculate the total flux in two ways: the first one is calculated from the original galaxy image ($F_{\text{Galaxy}}$) and the second one from the total model of the galaxy ($F_{\text{S\'ersic}}+F_{\text{PSF}}$). We compare the two slightly different AGN contribution fractions derived using GALFIT to better understand any possible bias introduced by fitting a S\'ersic model.

\begin{figure*}
    \centering
    \includegraphics[width=0.48
\textwidth]{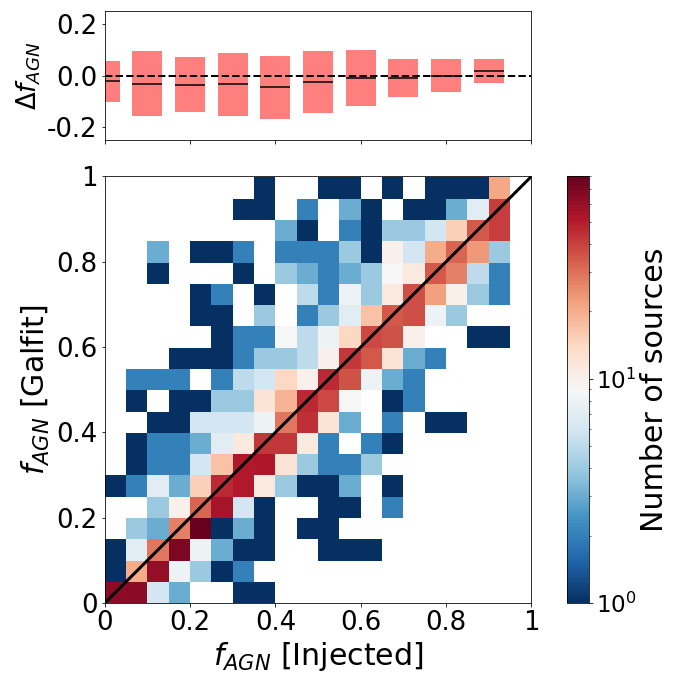}
    \includegraphics[width=0.48
\textwidth]{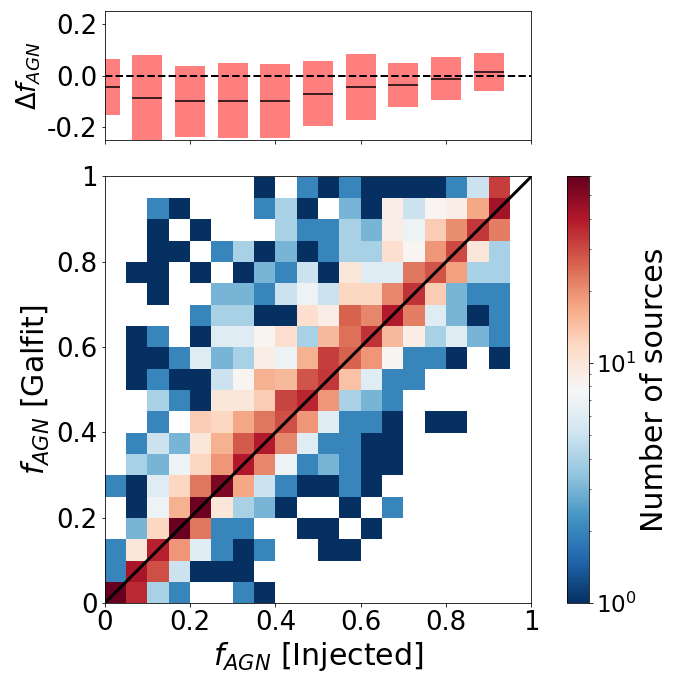}
    \caption{Comparison between the real injected AGN contribution fraction and the AGN contribution fraction obtained from GALFIT fitting, obtained by dividing the PSF flux by the total flux. In the left panel, the total flux corresponds to the flux measured within a 2\arcsec aperture in the original image, while in the right panel, the flux is measured within a 2\arcsec aperture in the model (S\'ersic+PSF). In both cases, the PSF flux is measured within the same aperture in the model PSF image. The dispersion in the left panel corresponds to $RMSE=0.12$, and in the right, $RMSE=0.17$. The colour bar provides the number of sources in each 2D bin.}
    \label{fig.predicted_AGNf_galfit}
\end{figure*}

In Fig.\,\ref{fig.predicted_AGNf_galfit} we compare the AGN contribution fraction obtained from running GALFIT (in the two ways explained above) with the real injected AGN contribution fraction. When we calculate the AGN contribution fraction by dividing by the total flux of the galaxy, we find that $\text{RMSE}=0.12$, with a mean difference (between predicted and real value) of $\langle{\Delta f_{AGN}}\rangle=-0.018$ and dispersion: $\sigma(\Delta f_{AGN})=0.12$. However, when we consider the AGN contribution fraction calculated completely from the model, we obtain slightly worse results, with $\text{RMSE}=0.17$, $\langle{\Delta f_{AGN}}\rangle=-0.075$ and $\sigma(\Delta f_{AGN})=0.15$. This highlights how an inaccurate S\'ersic model can bias the recovered AGN contribution fraction. From now on, we only show the results from calculating the total flux of the galaxy from the original galaxy image. The RMSE obtained from GALFIT is more than 10 times higher than that obtained from Zoobot. There seems to be a small systematic offset (as seen by the mean difference) that is not found in the Zoobot results, and a larger spread between predicted and real (injected) AGN contribution fractions than from Zoobot. In terms of dependence on galaxy structural properties, we find that galaxies with $n<1$, have an increased RMSE of 34\%, and it is even higher when the injected AGN contribution fraction is small ($f_{AGN}<0.2$), which means that GALFIT results are more affected by low S\'ersic index than our deep learning methods. We do not see a worse prediction of the $f_{AGN}$ for galaxies with high S\'ersic index for either method (see Fig.\,\ref{fig.apendix2} in Appendix \ref{sec.appendix.sersic} for more details on the effect of S\'ersic index in predicting $f_{AGN}$).

Figure \ref{rmse_z}, Fig.\,\ref{relative_error} and Fig.\,\ref{outlier_fraction_20} show the RMSE, relative error and outlier fractions, respectively, as defined in the previous section. The green lines show the results from the AGN contributino fraction obtained from the GALFIT fitting. We see similar trends to those from Zoobot, but with larger values (indicating significantly worse performance). We see that over all the galaxy properties ($f_{AGN}$, redshift, S/N and $r_{kron}$) considered, the RMSE is always higher for GALFIT than for Zoobot. Similar to the Zoobot results, we observe a small increase of RMSE with increasing redshift and with decreasing S/N. We also observe a general decrease in the RMSE with increased AGN contribution fractions, although the RMSE also decreases at $f_{AGN} < 0.1$. This decrease for very low AGN contribution fractions seems to arise due to the real AGN fraction value being very small. Actually, when looking at how the RAE changes with AGN contribution fraction (Fig.\,\ref{relative_error}) we see that it increases significantly as $f_{AGN}$ decreases. The same results are observed with Zoobot, but again, the total RAE for GALFIT is higher than for Zoobot, independently of $f_{AGN}$, redshift, S/N or $r_{kron}$. Consequently, the outlier fraction is also always higher for GALFIT, with a dramatic increase, which was also observed in the previous section, at $f_{AGN}<0.2$. 
In Table \ref{tab.results}, we summarise the results on the performance metrics obtained from GALFIT, for the two ways to calculate the AGN contribution fractions, that is, by either determining the total flux of the galaxy (within 2\arcsec) from the original galaxy or the GALFIT model (S\'ersic+PSF) and show the RMSE, RAE and outlier fractions (at different percentage levels, 20\% and 30\%) for the whole test sample.

\begin{figure*}
    \centering
    \includegraphics[width=0.48\textwidth]{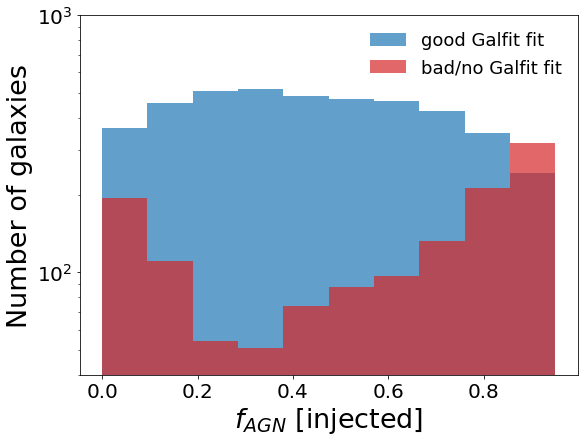}
    \includegraphics[width=0.48\textwidth]{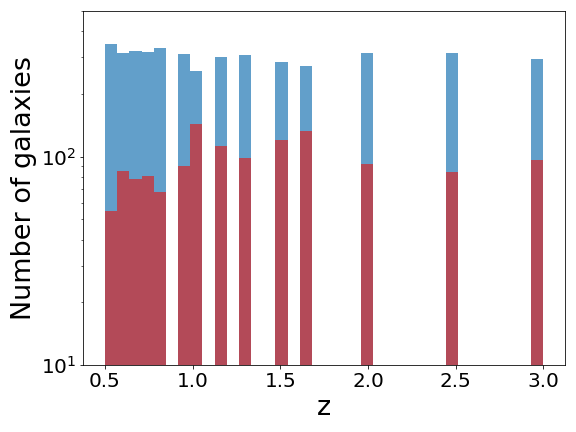}
    \includegraphics[width=0.48\textwidth]{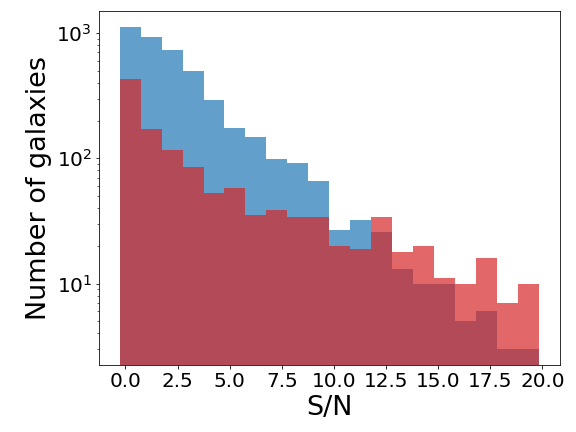}
    \includegraphics[width=0.48\textwidth]{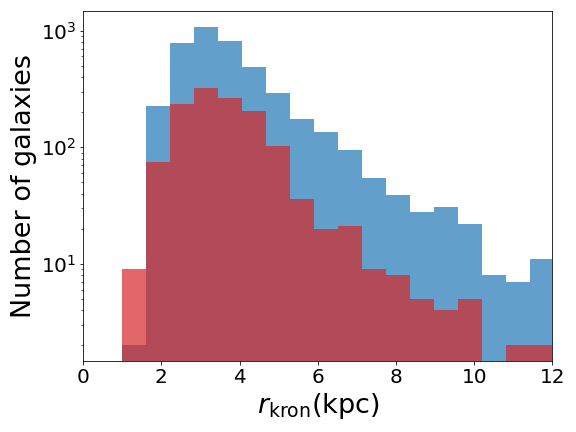}
    \caption{Number of galaxies for which GALFIT returns a good fit (blue histograms) and for which there is a bad fit or no fit from GALFIT (red histograms), as a function of the injected AGN contribution fraction (top left panel), redshift (top right panel), $S/N$ (bottom left panel) and $r_{kron}$ (bottom right panel). GALFIT tends to fail for bright galaxies, galaxies with low S/N, or with very high or low injected AGN contribution fractions.}
    \label{fig.bad_fits}
\end{figure*}

We find that $24\pm 1\%$ of all the galaxies from the subsample test set have no or bad fit from GALFIT. However, there is not a significant difference in the fraction of mergers and non-mergers that have bad or no fit. In Fig.\,\ref{fig.bad_fits}, we show that GALFIT fails more often for galaxies with high or low AGN contribution fraction (particularly at $f_{AGN}  \gtrsim 0.8$ and $f_{AGN}  \lesssim 0.2$), higher redshift, lower S/N and very small galaxies. Therefore, even though we have previously seen that both the RMSE and RAE are relatively low at $f_{AGN}  \gtrsim 0.8$, GALFIT fails to find a good fit in that regime for $46\pm 3\%$ of the galaxies, while at $f_{AGN}  \lesssim 0.2$ GALFIT fails for $25\pm 2\%$ of galaxies. GALFIT also fails more often at the higher redshifts ($28\pm 1\%$ of fails for $z>1$), compared to $19\pm 1\%$ for the rest of the sample, and $14\pm 2\%$ in the lowest redshift bin. Finally, GALFIT is more likely to fail for faint galaxies (with low S/N) and for very bright ones (with high S/N). There are $75\pm 3 \%$ of galaxies with no good fit for $S/N>8$ and  $24\pm 4 \%$ for $S/N<3$. These results highlight the importance of good-quality imaging in order to find a good fit to the galaxy's light with GALFIT, while our DL-based method can always determine a galaxy's AGN contribution fraction with good accuracy and precision regardless of these galaxy properties (within the dynamic ranges tested in this work).

\subsection{Application to \textit{JWST} galaxies}\label{sec.results.jwst}

\begin{figure}
    \centering
    \includegraphics[width=1\linewidth]{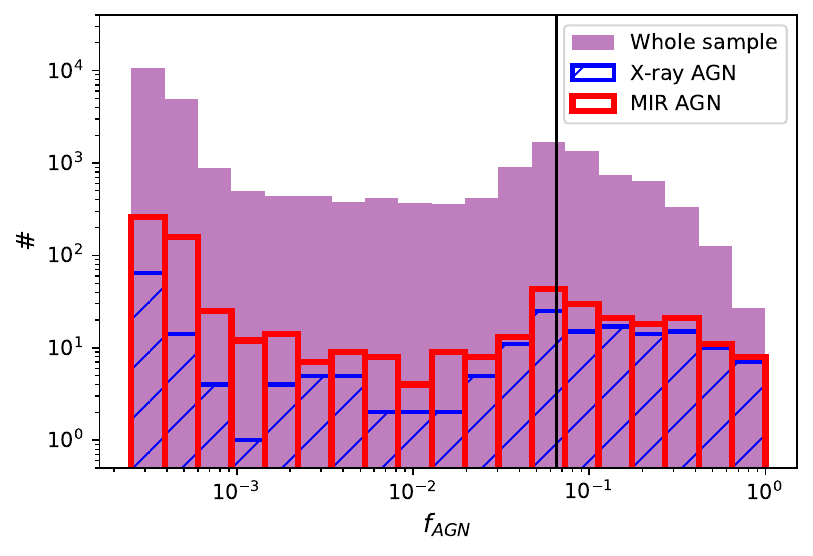}
    \caption{Distribution of the predicted $f_{AGN}$ on the sample of \textit{JWST} galaxies (purple). The blue and red histograms show the predictions for the X-ray, and MIR AGN within our sample, respectively. The black line ($f_{AGN}=0.065$) represents the 5$\sigma$ value with respect to the overall mean RMSE. The x- and y-axis are displayed in the logarithmic scale.}
    \label{fig.hist_fagn}
\end{figure}

Finally, we apply the trained DL model to our stellar mass complete sample of real \textit{JWST} galaxies as described in Section \ref{subsec.data.jwst}. In Fig.\,\ref{fig.hist_fagn}, we show the distribution of the DL predicted AGN contribution fractions. Considering the overal RMSE of our DL-based method in recovering the intrinsic $f_{AGN}$ is 0.013, a $5\sigma$ cut corresponding to $f_{AGN}=0.065$ can be used to select galaxies for which the method is confident in identifying a non-negligible contribution from a central point source component. Indeed, the distribution shown in Fig.\ref{fig.hist_fagn} clearly separates at around this $5\sigma$ cut value. 

It is important to note, however, that the AGN contribution fraction measures the nuclear light excess and may not necessarily be interpreted as exclusively tracing AGN emission, since compact star-forming regions, stellar clusters, or other unresolved components, particularly at high redshift, could also partially contribute to the detected central point-source signal. Additional multi-wavelength and/or spectroscopic diagnostics are therefore needed to confirm or refute the AGN nature of these candidates.

Many traditional methods of selecting AGN use a binary approach to separate galaxies into those which host AGN and those which do not. In reality, there is a continuous distribution of AGN contribution fraction. Therefore, although a binary AGN vs non-AGN selection approach is useful in identifying galaxies with dominant AGN, it is limited in the sense that it does not reflect the full spectrum. Adopting the $5\sigma$ cut at $f_{AGN}=0.065$, we find a total of $3337\substack{+218 \\ -196}$ galaxies hosting a non-negligile level of AGN activity in the 0.28 deg$^2$ of the COSMOS field analysed in this paper. The uncertainties correspond to varying the threshold by the RMSE of our method (0.013), i.e., counting galaxies with $f_{AGN}> 0.065 \pm 0.013$, which yields lower and upper thresholds of 0.052 and 0.078 respectively. Therefore, with respect to the total size of our stellar mass complete sample, 13\% of the galaxies are identified by our method to have a non-neglible level of AGN contribution in the {\it JWST}/NIRCam F150W filter.

Using the continuous parameter $f_{AGN}$ predicted by our DL method, users can choose different cuts on AGN contribution fraction to construct an AGN sample depending on the specific requirements. For example, we can classify galaxies as AGN based on the fractional AGN contribution, using a cut at $f_{AGN} > 0.2$, which leads to   $937\substack{+111 \\ -170}$
galaxies being classified as AGN over 0.28 deg$^2$. With a less conservative cut at $f_{AGN} > 0.1$, we find $2611\substack{+358 \\ -753}$ AGN within our sample. Furthermore, we find that there are 18 galaxies with $f_{AGN} > 0.7$ (shown in Fig.\,\ref{fig.example_dl7} in the appendix). There are five galaxies with spectroscopic information within those 18 galaxies. Four out of those five (i.e. 80\%) show optical broad lines in their spectra, as indicated by the specific flag in the release catalogue of spectroscopic redshifts available to the COSMOS collaboration from KECK follow up observations (Khostovan et al 2025, in prep). Based on this, we conclude that most galaxies with high $f_{AGN}$, as predicted by our method, are consistent with optical spectroscopically confirmed AGN (when that information is available).

To further investigate the nature of our $f_{AGN}$-selected galaxies, we perform a complementary analysis presented in Appendix \ref{sec.appendix.color}. Because the JWST/NIRCam F150W filter probes different rest-frame wavelengths across our redshift range, we train additional networks using the other three NIRCam filters available in COSMOS-Web (F115W, F277W and F444W) to partially mitigate redshift effect and provide additional colour information. In  Appendix \ref{subsec.appendix.analysis}, we examine the colours of AGN and non-AGN galaxies, as well as MIR-selected AGN. We find that our $f_{AGN}$-selected sources display systematically redder colours than non-AGN, in the same direction as the traditionally selected MIR AGN, which may be consistent with emission from dust heated by AGN.

Next, we compare our AGN based on the estimated $f_{\rm AGN}$ with AGN samples selected via X-ray detection or MIR colours.  Using the cut at $f_{\rm AGN}> 0.2$, we find that 20\,\% of the X-ray AGN are also selected as AGN based on the estimated $f_{\rm AGN}$, while 8\,\% of the MIR selected AGN are also selected by our method. If we use a less conservative cut at $f_{\rm AGN}> 0.1$, then the overlapping fractions increase for both AGN selections, to 15\,\% overlap with the MIR-selected AGN and 33\,\% overlap with the X-ray AGN. There is, however, a large fraction of the X-ray and MIR AGN with very low AGN contribution fraction $f_{\rm AGN}< 0.1$, and therefore, are not selected as AGN by our DL method based on a 0.1 threshold on $f_{AGN}$. This is to be expected, as it is well-known that some X-ray and MIR AGN are not detected by optical diagnostics \citep{Satyapal2008, Smith2014, LaMassa2019, Comerford2022}. This phenomenon could be explained by a bright host galaxy that dilutes the light of the AGN or by the presence of copious dust in the host galaxy, particularly when the galaxy is viewed edge-on, causing heavy obscuration \citep{Satyapal2008, Jackson2012, Smith2014, Hickox2018, Fitrina2022}. In Fig.\,\ref{fig.examples_xray_no_dl} and \ref{fig.examples_mir_no_dl} we show a random subset of the X-ray and MIR AGN with $f_{\rm AGN}< 0.065$ (i.e., below the $5\sigma$ cut based on the overall RMSE of our DL method), respectively. Many of these galaxies display spiral, clumpy or edge-on morphologies, in which higher dust content is expected. And in all cases, they do not show any obvious nuclear component in the {\it JWST}/NIRCAM F150W images, which explains the low predicted $f_{\rm AGN}$ despite being classified as X-ray or MIR AGN. \textbf{A similar analysis using the same methodology was done on Euclid/VIS imaging \citep{Margalef25}, where the overlaps with X-ray and MIR AGN selections were found to be higher. However, such variations are expected, as the resulting overlap fractions depend sensitively on several factors, including the intrinsic luminosity of the AGN population,  the rest-frame wavelength range probed by each dataset, and the specific selection criteria.}

We also computed the fraction of our AGN, based on $f_{\rm AGN}$ that are found in the X-ray and MIR-selected AGN samples. For sources with $f_{\rm AGN}> 0.2$, we find that 6\,\% are also X-ray AGN and 5\,\% are MIR AGN. These fractions decrease to 4\,\% and 3\,\%, respectively, when considering a lower threshold of 
$f_{\rm AGN}> 0.1$. This trend suggests that X-ray and MIR selection methods preferentially identify more powerful AGN, while our optical method is sensitive to a broader population, including less dominant or possibly obscured AGN. This result reinforces the notion that different AGN selection techniques probe different AGN populations. We also note that an $f_{\rm AGN}$ value above the selection threshold does not necessarily indicate true AGN activity; in some cases, the $f_{\rm AGN}$ signal may arise from central sources unrelated to AGN. We also investigate whether the $f_{\rm AGN}$  distributions differ for galaxies identified as X-ray or MIR AGN. As shown in Fig. \ref{fig.hist_fagn}, the distributions for these subsamples closely follow that of the whole sample, indicating that our method predicts similar $f_{\rm AGN}$  values regardless of whether the galaxy is independently classified as an AGN in X-ray or MIR. This suggests that the low overlap is not due to a bias in how our method treats X-ray or MIR AGN, but rather reflects the differences in the AGN populations each method is sensitive to.

\begin{figure*}
    \centering
    \includegraphics[width=1\linewidth]{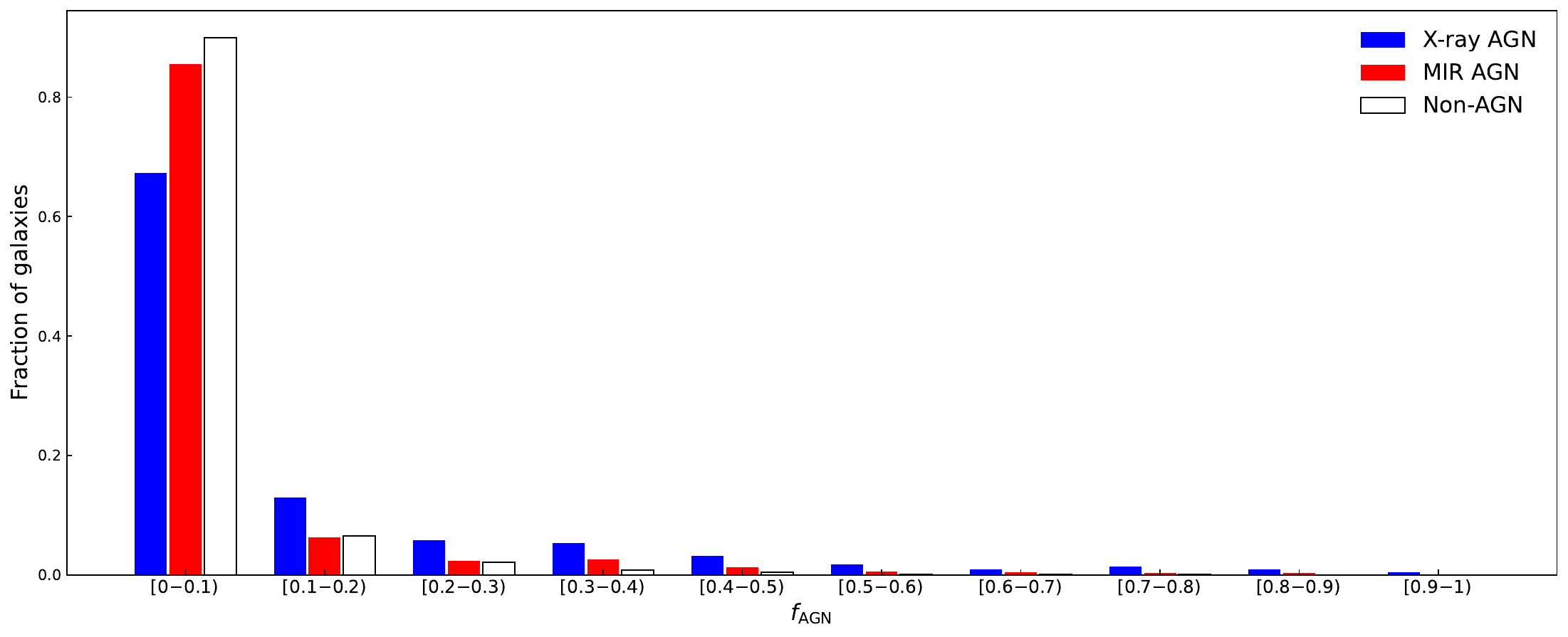}
    \caption{Fraction of galaxies in bins of predicted AGN contribution fraction in the X-ray AGN, MIR AGN and 'non-AGN' samples.}
    \label{fig.hist_agns}
\end{figure*}

In Fig.\,\ref{fig.hist_agns} we show the normalised distributions of the predicted $f_{\rm AGN}$ in the different AGN samples. The X-ray-selected AGN tend to dominated at higher AGN contribution fractions $f_{\rm AGN}>0.1$, indicating better correspondence with optically dominant AGN. For comparison, we also plot the distribution of the predicted $f_{\rm AGN}$ in the `non-AGN' sample which correspond to galaxies not identified as X-ray or MIR AGN in our sample. We can see that while a large fraction (around 90\,\%) of the non-AGN have $f_{\rm AGN}<0.1$, a small fraction (around 4\,\%) of them do have predicted $f_{\rm AGN}>0.2$, indicating that they can be AGN missed by the X-ray and MIR selections. In fact, it is widely known that the X-ray-selected AGN can miss a significant fraction of AGN with strong optical emission lines, particularly those that are heavily absorbed \citep{Heckman2005}, while the MIR selection can miss low-luminosity AGN or AGN in host galaxies dominated by starburst, or radio-loud AGN \citep{Hainline2016, Truebenbach2017}. Out of the 18 galaxies with $f_{AGN} > 0.7$ from our DL-based method as discussed previously, only a third of them are identified as MIR or X-ray AGN. However, we expect most of those 18 galaxies would be optically dominant AGN, based on the fact that five out of the 18 galaxies were followed up in optical spectroscopy and four exhibit broad lines in their spectra.

\begin{figure}
    \centering
    \includegraphics[width=1\linewidth]{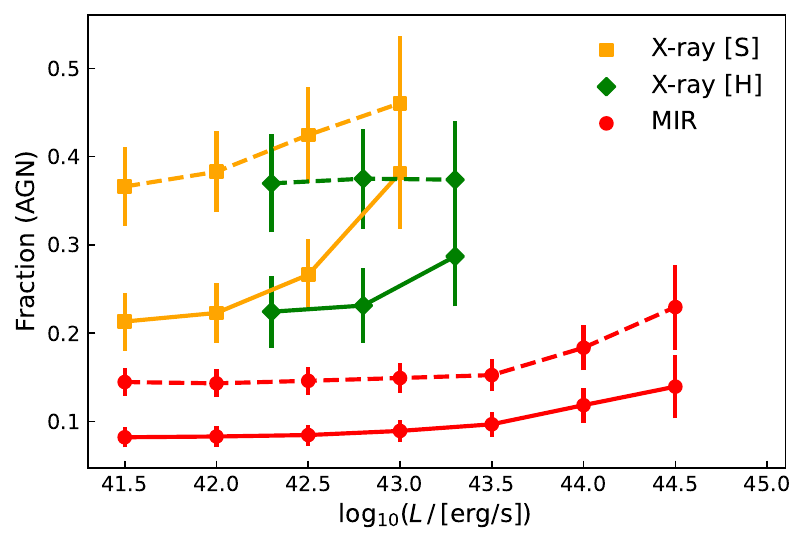}
    \includegraphics[width=1\linewidth]{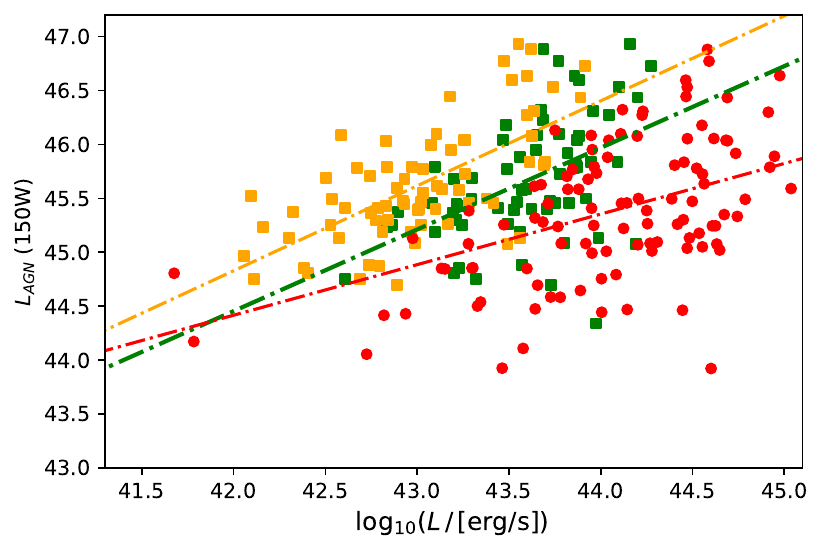}
    \caption{Top panel: Overlapping fraction of AGN also identified by our DL method as a function of the adopted cut on the X-ray or the rest-frame 6 $\mu$m luminosity for the X-ray and MIR-selected AGN. The solid lines show the overlapping fraction by applying a cut at $f_{\rm AGN}> 0.2$, while the dashed lines show the overlapping fraction if we select AGN by requiring $f_{\rm AGN} > 0.1$. Bottom panel: AGN luminosity in the {\it JWST}/NIRCAM F150W filter versus the X-ray luminosity or the rest-frame 6 $\mu m$ luminosity, for the X-ray-selected or MIR-selected AGN. The dash-dotted lines show the respective linear fits.}
    \label{fig.frac_lum_xray}
\end{figure}

The top panel of Fig.\,\ref{fig.frac_lum_xray} shows the overlapping fraction of AGN also identified using our DL method based on the predicted AGN contribution fraction as a function of the adopted cut on the X-ray luminosity $L_X$ or the rest-frame 6 $\mu m$ luminosity $L_{6 \mu m}$, for the X-ray and MIR selected AGN, respectively. We see that the overlapping fraction of AGN also selected by our model increases with increasing luminosity (more slowly for the MIR AGN). This is expected as the X-ray luminosity and the rest-frame 6 $\mu m$ luminosity (which are indicators of the power of the X-ray or MIR-selected AGN) correlate, albeit with significant scatter, with the luminosity of the AGN component in the {\it JWST}/NIRCAM F150W filter $L_{\rm AGN}$  (calculated as the total luminosity in that filter multiplied by $f_{\rm AGN}$). We show these correlations in the bottom panel of Fig.\,\ref{fig.frac_lum_xray}, only for galaxies with $f_{\rm AGN}>0.065$ for which we can confidently identify a central point source contribution). The strongest correlation is found for the X-ray AGN detected in the soft band, and the weakest correlation is found in the MIR AGN. 

\section{Summary and Conclusions}\label{sect:Conclusions}

In this paper, we presented a new DL-based method Zoobot to determine the AGN contribution fraction to the total flux of a galaxy, trained on realistic mock images generated from cosmological hydrodynamical simulations. Specifically, we constructed the training sample for our DL model from the IllustrisTNG simulations. First, we mimicked the observational effects of the\textit{JWST}/COSMOS-WEB survey in the NIRCAM F150W filter, by using the same pixel scale, convolving with the full range of the survey PSFs, adding Poisson noise and finally, adding the real \textit{JWST} sky background. Then, we artificially added an AGN component (approximated by the PSF) at different levels to represent different AGN contribution fractions to the total flux. After training our model we analysed the performance results in detail on a test sample. Furthermore, we used GALFIT to fit the 2D light profile of the galaxies in the test sample with a combination of S\'ersic + PSF profiles to calculate the AGN contribution fraction within a circular aperture. We showed how our DL method outperforms the traditional 2D surface brightness fitting method and is 2500 times faster in inferring the AGN fraction. Our main results are summarised in the following: 

\begin{enumerate}
    \renewcommand{\labelenumi}{\it \roman{enumi})}
    
    \item The AGN contribution fraction $f_{AGN}$ predicted by Zoobot has a very tight correlation with the injected $f_{AGN}$, with a mean difference of $-0.0018$ and an overall RMSE $=0.013$ for the whole test sample over the redshift range $0.5<z<3$, demonstrating how well our DL method is at recovering the true AGN contribution to the total flux. The RMSE performance metric is around 10 times better than that obtained using GALFIT. In both methods, there is an increase in the RMSE at lower  $f_{AGN}$ values, higher redshifts, lower S/N and smaller galaxy sizes. 
    
    \item The outlier fraction given by Zoobot, as measured by the number of galaxies that have RAE $>20\%$, is below 8\% for most of the sample, except for galaxies with very low AGN contribution fractions ($f_{AGN}<0.1$) and faint galaxies (mag $>22$ in the F150W filter). There is only a small increase with lower S/N. However, the outlier fraction is most affected by the AGN contribution fractions and starts to increase sharply at $f_{AGN}<0.2$. The outlier fraction obtained from the GALFIT results is on average around 20\%, and similarly to our Zoobot results, there is an increase with decreasing $f_{AGN}$ and S/N. Furthermore, there is a significant increase in the outlier fraction at higher redshift, which is not observed in the Zoobot results.
    
    \item In constructing the training sample for our DL model, we fully incorporated the temporal and spatial variations of the PSF, which helps our model to better determine the intrinsic AGN contribution fraction. 
    At $f_{AGN}>0.2$, the precision of our DL method is even better than the level of intrinsic variations in the observed PSF.
    Once it is trained, our DL model can be easily applied to new data, without the need for user inputs as in traditional 2D fitting codes (such as GALFIT) do. Furthermore, it can output $f_{AGN}$ predictions of thousands of galaxies in a few seconds, making it an ideal method for large extra-galactic imaging surveys. 
    \item We applied the trained DL model to real \textit{JWST} observations, and we find that $937\substack{+111 \\ -170}$ have  $f_{AGN} > 0.2$, while $2611\substack{+358 \\ -753}$ have $f_{AGN} > 0.1$. Moreover, when comparing with other AGN selection methods, we find that 20\% of the X-ray AGN and 8\% of the MIR selected AGN are also selected as AGN based on the selection of $f_{\rm AGN}>0.2$. Using the selection of $f_{\rm AGN}>0.1$, the overlapping fractions increase to 33\% and 15\% for X-ray and MIR AGN, respectively. 
\end{enumerate}

In future works, we will extend the comparison of our deep learning predictions with other AGN selection techniques, such as those identified using optical spectroscopy (type 1 and type 2 optical AGN) and radio observations. A follow-up and complementary approach that we also aim to develop in the near future will focus on not only inferring the level of AGN contribution fraction to the observed total light but also decomposing the observed images into a pure AGN component and a host galaxy component. This will allow us to investigate host galaxies of AGN in more detail. For example, we can look for the presence of merging signs and bar features which could trigger AGN.

\section*{Acknowledgements}

This publication is part of the project `Clash of the Titans: deciphering the enigmatic role of cosmic collisions' (with project number VI.Vidi.193.113 of the research programme Vidi which is (partly) financed by the Dutch Research Council (NWO). We thank Mara Salvato for helpful discussions on AGN. We thank the Center for Information Technology of the University of Groningen for their support and for providing access to the Hábrók high-performance computing cluster. We thank SURF (www.surf.nl) for the support in using the National Supercomputer Snellius. 

\bibliography{references}

\begin{appendix}
 \onecolumn

\section{Results as a function of S\'ersic index}\label{sec.appendix.sersic}
\begin{figure*}[h]
    \centering
    \includegraphics[width=0.48\textwidth]{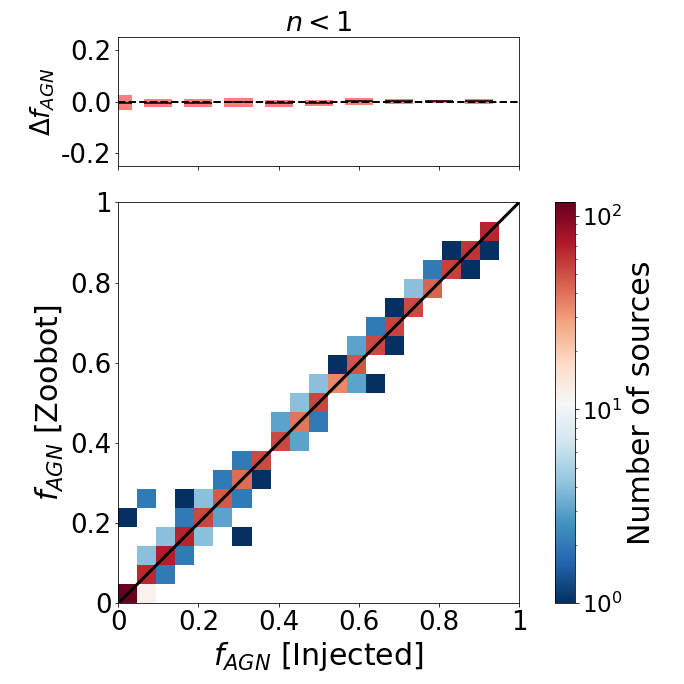}
    \includegraphics[width=0.48\textwidth]{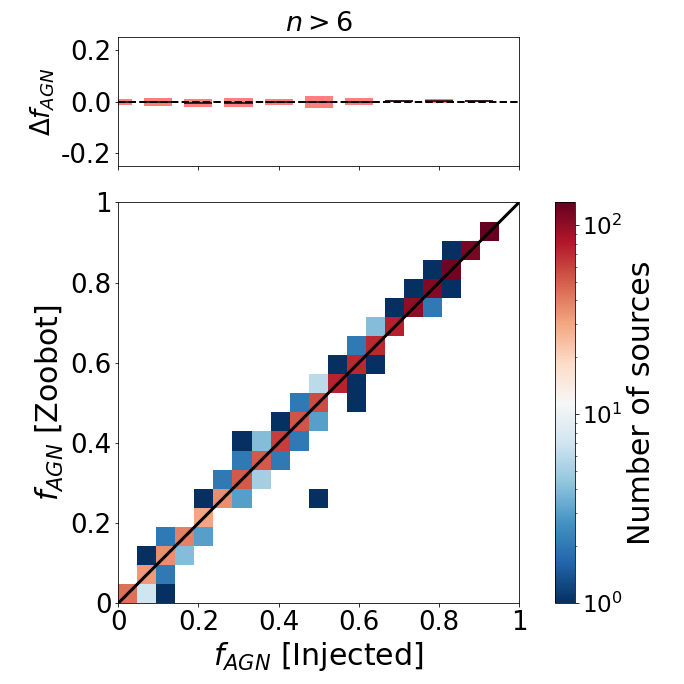}\vspace{3ex}
    \includegraphics[width=0.48\textwidth]{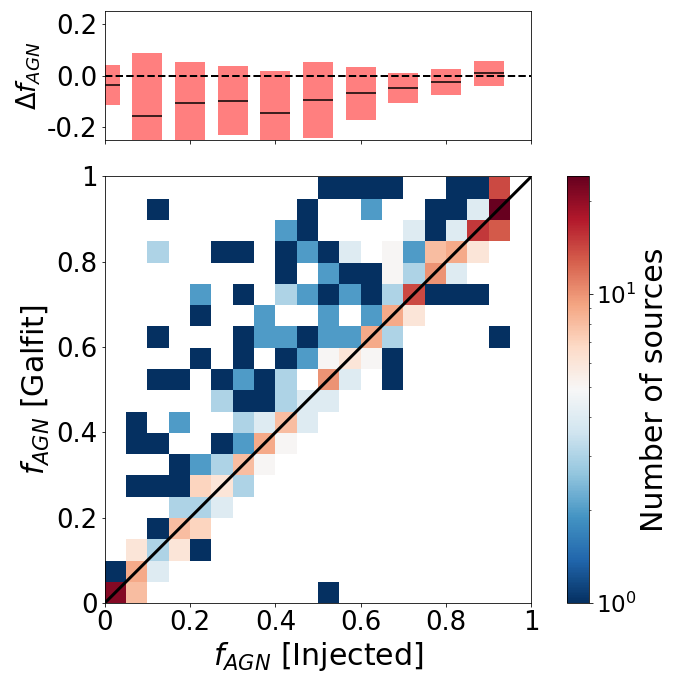}
    \includegraphics[width=0.48\textwidth]{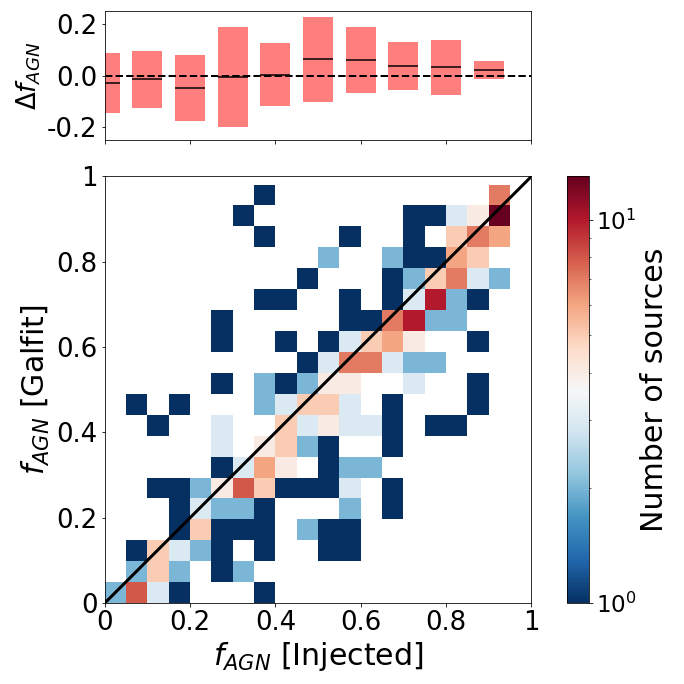}
    \caption{Comparison between the real injected AGN contribution fraction and the recovered AGN contribution fraction obtained from the Zoobot (top row) and GALFIT (bottom row), for galaxies with low S\'ersic index ($n<1$, left panels) and high S\'ersic index ($n>6$, right panels). The colour bar indicates the number of sources in each 2D bin.}
    \label{fig.apendix1}
\end{figure*}

\begin{figure}
    \centering
    \includegraphics[width=0.33\textwidth]{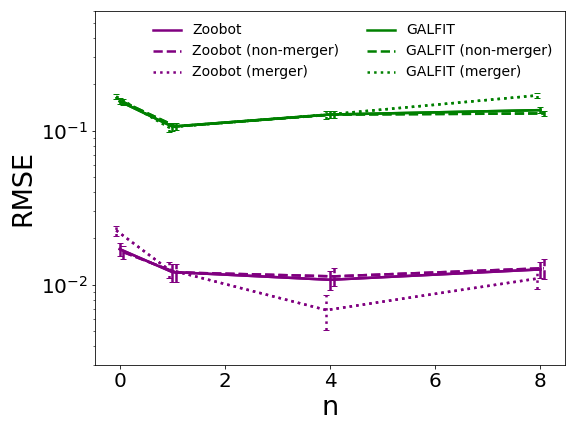}
    \includegraphics[width=0.33\textwidth]{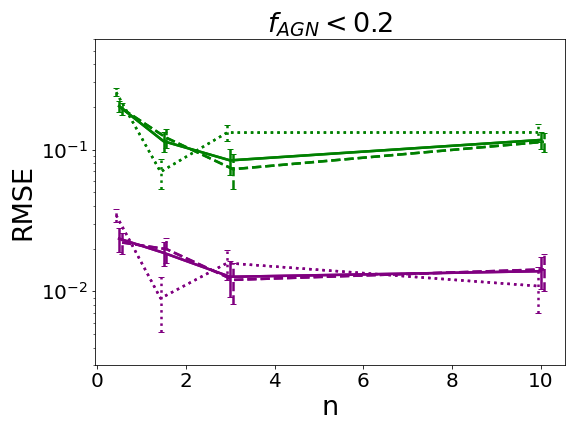}
    \includegraphics[width=0.33\textwidth]{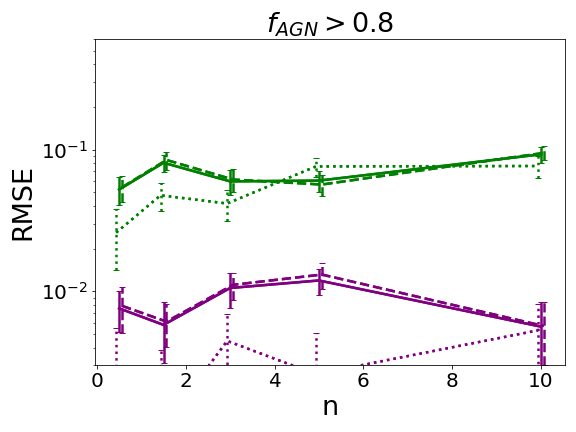}
    \caption{RMSE as a function of S\'ersic index, for the whole sample (top panel), for galaxies with $f_{AGN}<0.2$ (centre panel) and galaxies with $f_{AGN}>0.8$ (bottom panel). The purple lines correspond to the results from Zoobot and the green lines from GALFIT. The solid lines correspond to the whole sample while the dashed and dotted lines correspond to the mergers and non-merger galaxies, respectively.}
    \label{fig.apendix2}
\end{figure}

We analyse how the performance of both methods depends on S\'ersic index. In Fig.\,\ref{fig.apendix1} we compare the predicted $f_{AGN}$ values from Zoobot and GALFIT for galaxies with low S\'ersic index ($n<1$) on the top panels and galaxies with high S\'ersic index ($n>6$) on the bottom panels. The results from Zoobot for low S\'ersic index galaxies have $\text{RMSE}=0.017$, $\langle{\Delta f_{AGN}}\rangle=-0.0025$ and $\sigma(\Delta f_{AGN})=0.017$, which corresponds to an increase of 31\% in the RMSE. For high S\'ersic galaxies, the performance is slightly better than the average for all galaxies, with $\text{RMSE}=0.013$, $\langle{\Delta f_{AGN}}\rangle=-0.0007$ and $\sigma(\Delta f_{AGN})=0.012$. On the other hand, low S\'ersic index galaxies have, from GALFIT results, a 47\% worse RMSE compared to the whole test sample, with $\text{RMSE}=0.16$, $\langle{\Delta f_{AGN}}\rangle=-0.073$ and $\sigma(\Delta f_{AGN})=0.14$, and it appears that the results are worse when not only the S\'ersic index is low, but also the $f_{AGN}$ (this is also seen in Fig.\,\ref{fig.apendix2}), where we show how the RMSE varies in different bins of $n$ for the whole sample, but also for galaxies with low and high AGN fraction ($f_{AGN}<0.2$ and $f_{AGN}>0.8$, respectively). A high S\'ersic index does not seem to impact the predictions of $f_{AGN}$ determined from GALFIT, with $\text{RMSE}=0.094$, $\langle{\Delta f_{AGN}}\rangle=-0.005$ and $\sigma(\Delta f_{AGN})=0.094$.

\section{AGN example images}\label{sec.appendix.examples}

\begin{figure*}[h]
    \includegraphics[width=1\textwidth]{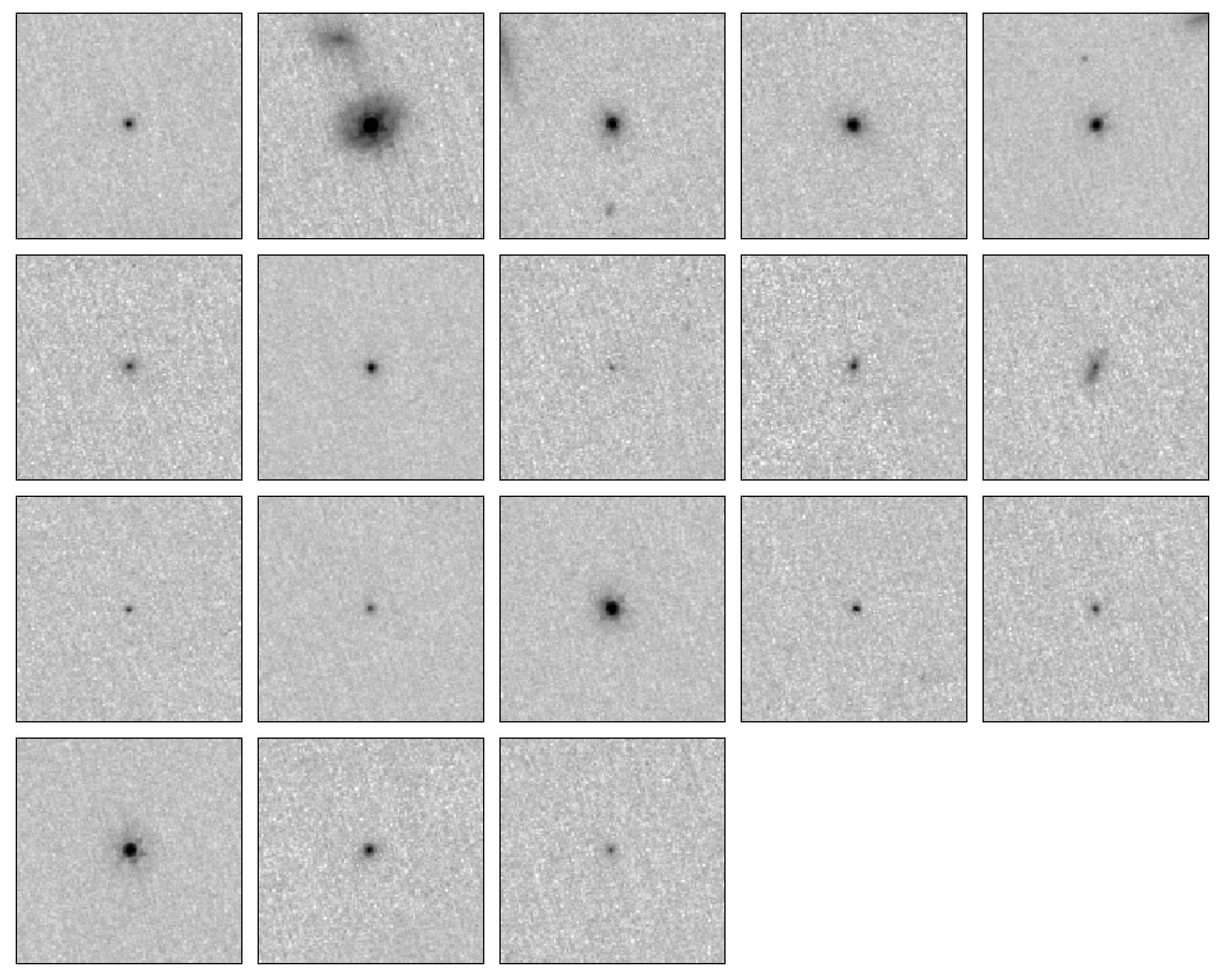}
    \caption{Example galaxies with $f_{AGN}>0.7$. Images are 3.84\arcsec across and are displayed with an inverse arcsinh scaling.}\label{fig.example_dl7}
\end{figure*}

\begin{figure*}
    \includegraphics[width=1\textwidth]{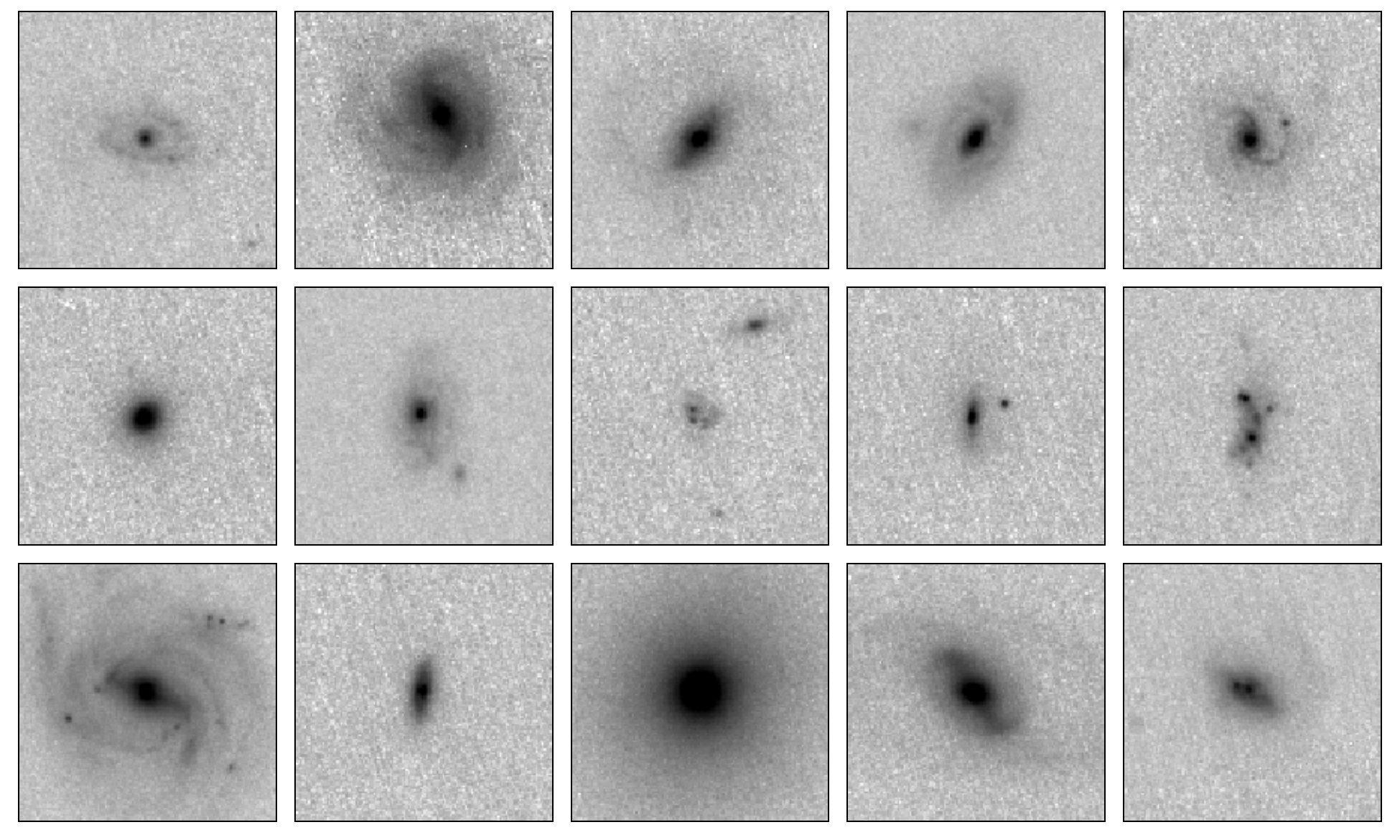}
    \caption{Example of X-ray AGN with $f_{AGN}<0.065$. Images are 3.84\arcsec across and are displayed with an inverse arcsinh scaling.}\label{fig.examples_xray_no_dl}
\end{figure*}

\begin{figure*}
    \includegraphics[width=1\textwidth]{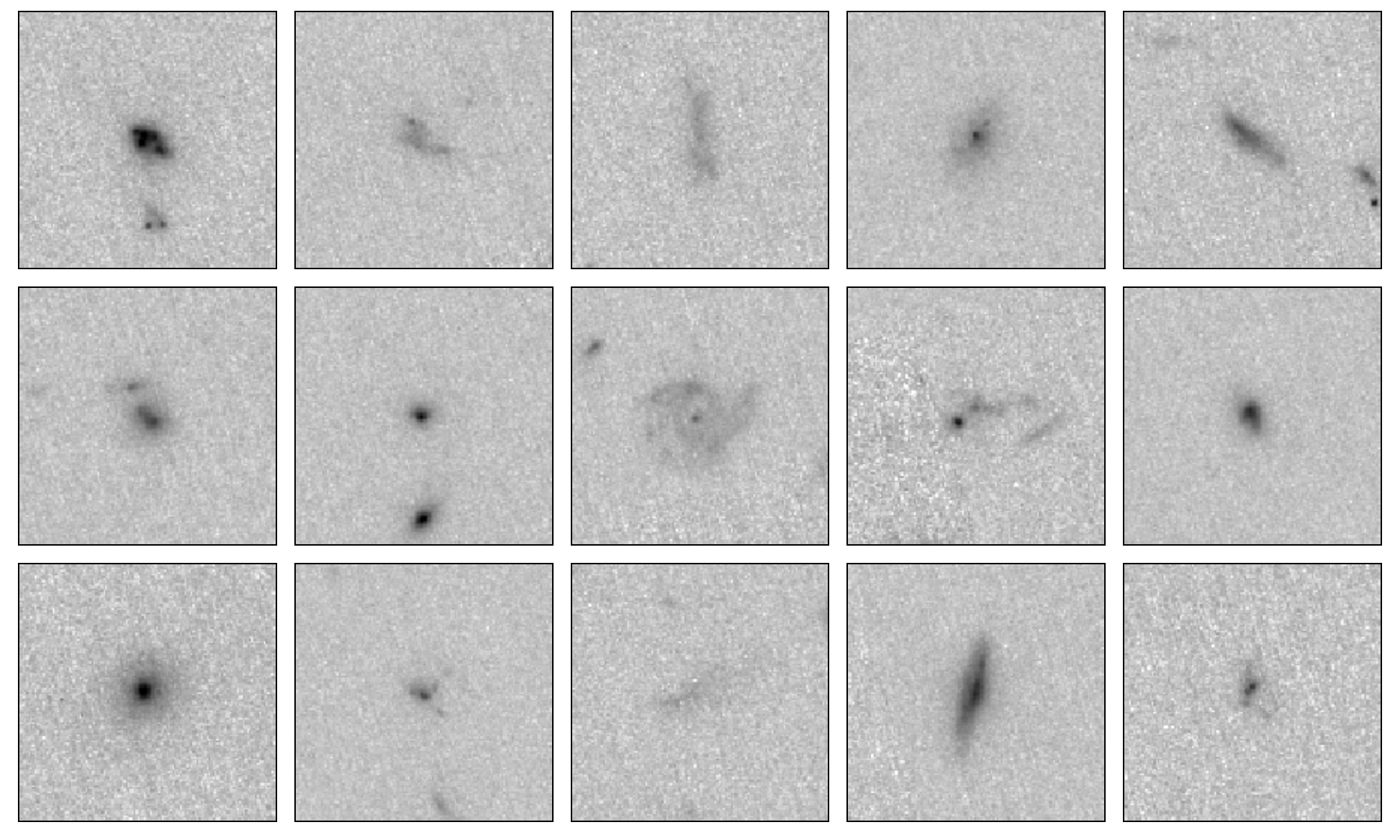}
    \caption{Example of MIR AGN with $f_{AGN}<0.065$. Images are 3.84\arcsec across and are displayed with an inverse arcsinh scaling.}\label{fig.examples_mir_no_dl}
\end{figure*}

We show in Fig.\,\ref{fig.example_dl7} the 18 galaxies in our sample for which our DL model predicts $f_{\rm AGN} > 0.7$. This galaxies display a bright central source. In Fig.\,\ref{fig.examples_xray_no_dl} and \ref{fig.examples_mir_no_dl} we show random subsets of X-ray and MIR AGN, respectively, with very low $f_{\rm AGN}$ predicted by the model. These galaxies do not show a central component in the F150W filter.

\section{Color-color selection}\label{sec.appendix.color}

\subsection{Training in Multiple JWST Filters}

We retrained our deep learning model on mock data constructed for the additional NIRCam filters: F115W, F277W, and F444W. The mock data for these filters were generated analogously to F150W (see Sect. \ref{subsec.data.mock}), with the only differences being the use of the specific filter response curves and PSF models for each band, and the use of cutouts corresponding to the respective filter. The procedure for creating model AGN-host galaxies is identical to that described for F150W (see Sect. \ref{sect:mock.agn}). The training procedure followed the same setup as for F150W (see Sect. \ref{sect:CNN} of the main text), employing the same ConvNeXt-Base architecture pretrained on Zoobot for each dataset.

Table \ref{tab.results.models} summarises the model performance in each filter, reporting the RMSE, RAE, mean difference, and outlier fractions (at 20\% and 30\% thresholds) in the recovery of $f_{AGN}$ across the test sets. The performance is comparable across all filters, indicating that the model is robust to band choice.

\begin{table*}[h]
\caption{Zoobot model performance across JWST filters.}
    \centering
    \begin{tabular}{|l|c|c|c|c|c|c}
    \hline
                &  RMSE             & RAE & $\langle{\Delta f_{AGN}}\rangle$ & Outlier (20\%)    & Outlier (30\%)    \\
               \hline
        F150W   & $0.013 \pm 0.005$ & $0.076 \pm 0.005$ & $-0.0018 \pm 0.0002$ & $0.065 \pm 0.004$ & $0.061 \pm 0.003$ \\
        F115W   & $0.046 \pm 0.006$ & $1.14 \pm 0.04$ & $0.0033 \pm 0.0002$ & $0.147 \pm 0.004$ & $0.084 \pm 0.003$ \\
        F277W   & $0.026 \pm 0.002$ & $0.16 \pm 0.02$ & $0.0034 \pm 0.0001$ & $0.110 \pm 0.010$ & $0.079 \pm 0.010$ \\      
        F444W   & $0.029 \pm 0.002$ & $0.83 \pm 0.03$ & $0.0047 \pm 0.0001$ & $0.140 \pm 0.004$ & $0.104 \pm 0.004$ \\  
        \hline
    \end{tabular}
    \tablefoot{We summarise the overall performance from Zoobot, in terms of the RMSE, RAE, mean difference and outlier fractions (at different percentage levels, 20\% and 30\%). }
    \label{tab.results.models}
\end{table*}

\subsection{Application to Real JWST Data and AGN Flux Calculation}\label{subsec.appendix.jwst}

We applied the trained models to the real JWST/NIRCam data in each filter (F115W, F150W, F277W, and F444W) for the same galaxy sample described in the main text, obtaining a prediction of $f_{AGN}$ in each band. To compute the AGN fluxes, we require total galaxy fluxes, which we adopt from Chen et al. (2025, in prep). Briefly, Chen et al. (2025, in prep) used JWST/NIRCam images reduced by \cite{Zhuang2024}. The images were processed using JWST pipeline v1.10.2 with CRDS v11.17.0, along with additional custom steps. Source photometry and morphology were derived by modelling each galaxy with a single S\'ersic profile using GALFIT \citep{Peng2002}, with initial parameters from SExtractor \citep{Bertin1996}. This process provides both geometric (e.g., axis ratio, position angle) and physical (e.g., flux, magnitude, S\'ersic index, effective radius) properties.

The resulting JWST/NIRCam magnitudes were cross-validated against the nearest bands in COSMOS2020 (F115W vs. VISTA NB118, F150W vs. VISTA H, F277W vs. VISTA Ks, and F444W vs. IRAC CH2), showing tight correlations and good agreement. The total fluxes are then combined with the $f_{AGN}$ values from our models to derive the nuclear (AGN) fluxes in each band.

\subsection{Multi-Band Analysis}\label{subsec.appendix.analysis}

\begin{figure}
    \centering
    \includegraphics[width=1\linewidth]{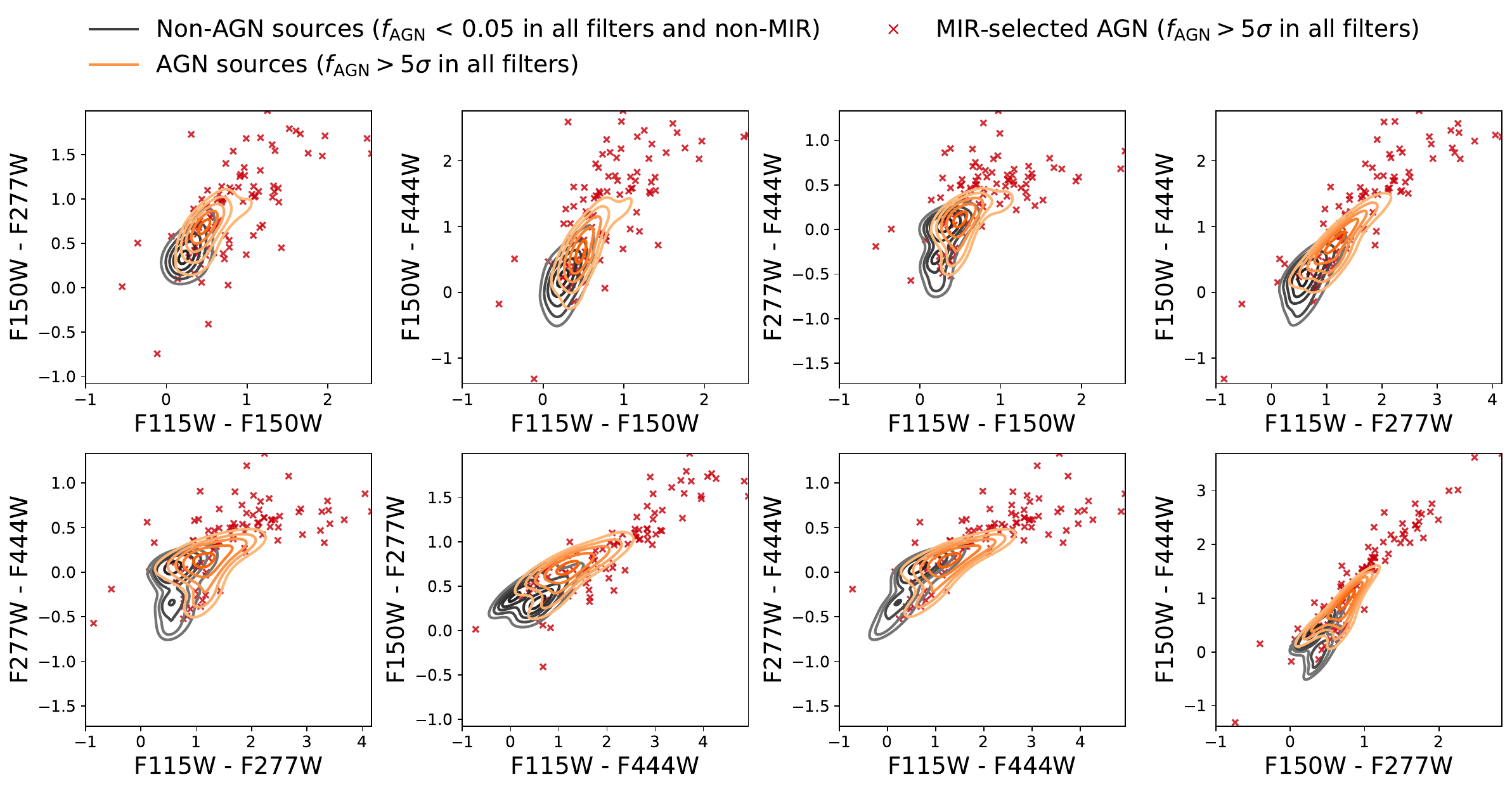}
    \caption{Colour–colour diagrams for eight representative filter combinations from the four JWST/NIRCam bands (F115W, F150W, F277W, F444W). Black contours show the distribution of non-AGN galaxies ($f_{AGN}<0.05$ in all bands). Orange contours indicate the $f_{AGN}$-selected AGN based on the whole-galaxy fluxes. MIR-selected AGN are overplotted as red crosses (if they satisfy $f_{AGN}>5\sigma$).}
    \label{fig:color-color-nuclei}
\end{figure}

Using the AGN fractions derived in each JWST/NIRCam filter, we performed a multi-band analysis to assess the nature of the central light excess, by constructing color–color diagrams in various filter combinations (e.g., F150W–F277W vs. F115W–F150W). In Figure \ref{fig:color-color-nuclei} we show a subsample of colour–colour diagrams constructed from the total galaxy fluxes in the four JWST/NIRCam filters.

\begin{figure}
    \centering
    \includegraphics[width=1\linewidth]{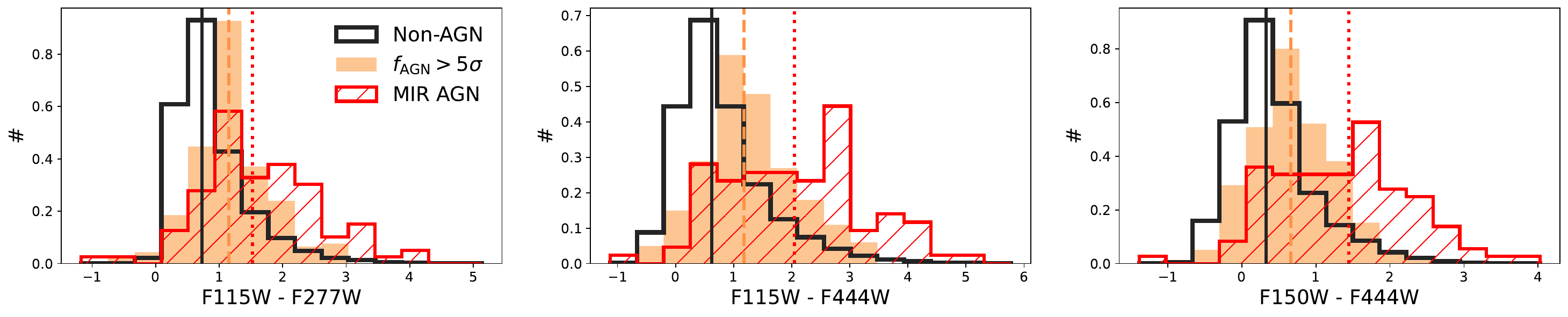}
    \caption{Normalised colour distributions of our galaxy sample (F115W-F227W, F115W-F444W, and  F150W-F444W). Galaxies with negligible nuclear light ($f_{AGN}<0.05$ in all filters and not selected as MIR AGN) are shown in black, AGN candidates with significant nuclear light ($f_{AGN}>5\sigma$ in all bands) in orange, and MIR-selected AGN in red. Vertical lines indicate the median magnitude of each distribution.}
    \label{fig:color-color-histogram}
\end{figure}

In each diagram, we show the contours of the non-AGN population (black), defined as galaxies with  $f_{AGN}<0.05$ in all four filters. The AGN sample is defined as galaxies with $f_{AGN}>5\sigma$ in each filter (orange), i.e. systems with a measurable AGN contribution across all bands. We further overplot the MIR-selected AGN (red), for which $f_{AGN}>5\sigma$ in all filters. We focus on this representative subsample of colour–colour combinations because they most clearly highlight the systematic differences between the AGN and non-AGN populations. The $f_{AGN}$-selected AGN are, on average, redder than the non-AGN galaxies, which may suggest that their host galaxies are more evolved and less actively star-forming, and thus less consistent with young star-forming regions. This reddening could also be driven by the presence of hot dust heated by the AGN, which contributes additional emission at longer wavelengths. The MIR-selected AGN span a broader range of colours but largely overlap with the$f_{AGN}$-selected AGN, often extending toward even redder colours, possibly indicating even higher dust temperatures or more substantial dust obscuration in these systems. To highlight the differences between the non-AGN population and the AGN sample (defined here as galaxies with $f_{AGN}>5\sigma$ in each filter), we show, in Figure \ref{fig:color-color-histogram} the three colours with the largest median differences between the two populations. These plots also illustrate that the MIR-selected galaxies exhibit more extreme (redder) colours, while our selection tends to identify AGN whose emission is less distinct from the stellar component, as reflected by their more moderate colours.

By performing the $f_{AGN}$ analysis across multiple JWST/NIRCam filters, we can partially mitigate the rest-frame wavelength differences introduced by the wide redshift range of our sample ($0.5<z<3$). While any single filter probes different physical regimes at different redshifts, the combination of multiple bands allows us to compare nuclear light excesses more consistently across the sample. This approach reduces, though does not fully eliminate, potential biases due to varying SED coverage. A more detailed analysis accounting explicitly for rest-frame band differences and nuclear versus host SEDs will be required to fully quantify these effects.

\end{appendix}

\end{document}